\documentclass[%
reprint, 
twocolumn,
superscriptaddress,
amsmath,amssymb,
aps,
prd,
floatfix,
]{revtex4-2}

\usepackage{graphicx}               
\usepackage[export]{adjustbox}      
\usepackage{dcolumn}                
\usepackage{bm}                       
\usepackage{xcolor}                 
\usepackage{amsmath}                
\usepackage{amssymb}                 
\usepackage{braket}                 
\usepackage{booktabs}
\usepackage[hidelinks]{hyperref}
\usepackage[T1]{fontenc}
\usepackage{microtype}

\begin{document}

\preprint{}

\title{Transport Properties of the MRI in Differentially Rotating Neutron Stars}

\author{Harshraj Bandyopadhyay}
\email{hkb5238@psu.edu}
\affiliation{Institute for Gravitation and the Cosmos, The Pennsylvania State University, University Park, PA 16802, USA}
\affiliation{Department of Physics, The Pennsylvania State University, University Park, PA 16802, USA}
\author{Peter Hammond}
\affiliation{Max Planck Institute for Gravitational Physics (Albert Einstein Institute), 14476 Potsdam, Germany}
\author{David Radice}
\affiliation{Institute for Gravitation and the Cosmos, The Pennsylvania State University, University Park, PA 16802, USA}
\affiliation{Department of Physics, The Pennsylvania State University, University Park, PA 16802, USA}
\affiliation{Department of Astronomy \& Astrophysics, The Pennsylvania State University, University Park, PA 16802, USA}

\date{\today}

\begin{abstract}
We perform three-dimensional, high-resolution, general-relativistic magnetohydrodynamics (GRMHD) simulations of the magnetorotational instability (MRI) in differentially rotating neutron stars. We consider both high-mass models which collapse either promptly, or due to the outward transport of angular momentum removing rotational support, and lower-mass models, which remain stable even after solid-body rotation has been achieved. We measure effective transport coefficients of the resulting turbulent flow, shear-viscosity $\alpha_{\rm vis}$, mixing-length $\ell_{\rm mix}$, and mean-field dynamo $\alpha_{\rm DYN}$-parameter, and their correlations with mean-flow parameters, such as rest-mass density $\rho$, characteristic wavelength of the MRI $\lambda_{\rm MRI}$, and vertical Alfv\'en velocity $v_A$. At saturation, $\ell_{\rm mix}$ and $\alpha_{\rm DYN}$ correlate with $\lambda_{\rm MRI}$ and $v_A$ respectively, as expected from dimensional analysis. However, both quantities show deviations of roughly an order of magnitude from the values predicted on the basis of these correlations. They are also highly intermittent in space and time. The mixing length is found to be largely independent of density, calling into question the $\ell_{\rm mix}(\rho)$ ansatz used in many general-relativistic large-eddy simulations. Our results show that, while the effects of MRI-induced turbulence might be qualitatively reproduced by simulations that employ transport coefficients chosen using dimensional-analysis considerations, fully-resolved GRMHD simulations are needed to make quantitative predictions.
\end{abstract}

\maketitle

\section{Introduction}
\label{sec:introduction}
Binary neutron star (BNS) mergers and core-collapse supernovae (CCSNe) are among the most important laboratories in the universe where we can observe some of the most magnetically active phenomena. The joint detection of GW170817 with its short gamma-ray burst (SGRB) \cite{Abbott2017GW170817,Abbott2017GWGRB} and kilonova \cite{Kasen2017Origin,Metzger2020Kilonovae} plays an important role in understanding both these cosmic events. Each of these events produces a hot, differentially rotating compact object in the form of a hypermassive or supramassive neutron star (HMNS/SMNS) following merger \cite{ShibataUryu2000,Baumgarte2000MaximumMass}, 
or a proto-neutron star (PNS) embedded in a rapidly rotating envelope after stellar collapse \cite{Janka2007CoreCollapse}. 
In both of these settings, the magnetorotational instability (MRI) plays a crucial role in the evolution of these systems. The MRI can amplify weak seed magnetic fields by a factor of $10-100$~\cite{rembiasz2016maximum}. It drives the differential rotation redistribution and creation of large-scale magnetic structure required to launch a relativistic jet \cite{Duez2006Evolution,Kiuchi2015KelvinHelmholtz,Kiuchi2018Global,KiuchiEtAl2024,Kiuchi2026Magnetar,MostQuataert2023,CombiSiegel2023Jets,Musolino2025Neutrinos}. 

The MRI is a local, linear instability of weakly magnetized, differentially rotating fluids \cite{BalbusHawley1991}. A fluid element displaced radially is connected to its neighboring element by magnetic tension and this tension allows the elements to exchange angular momentum. When the angular velocity $\Omega$, decreases outward, $\partial_\varpi \Omega<0$, the momentum exchange destabilizes the system and perturbation grows exponentially. Here $\varpi$ is the cylindrical radius, and $\Omega = u^\phi/u^t$, with $u^\alpha$ as the 4-velocity of the fluid element. \cite{BalbusHawley1991,BalbusHawley1998}. As the MRI reaches saturation it produces magnetohydrodynamic (MHD) turbulence, and, in this turbulent state, the Maxwell stress, $M_{\varpi\phi}=\langle -b_{\varpi}b_{\phi}\rangle$, transports the angular momentum outward much more efficiently than microscopic processes like kinetic viscosity \cite{BalbusHawley1998,Balbus2003EnhancedTransport}. As the MRI grows rapidly (on a timescale comparable to the rotation period) and generates dynamically strong magnetic fields, it becomes an important mechanism for magnetic-field evolution in sufficiently ionized, differentially rotating gas in the ideal-MHD regime. 
To understand the gravitational-wave signatures, the electromagnetic emission, and the heavy-element yield of the ejecta of these cosmic events, we need a quantitative understanding of MRI-driven turbulence in compact-object interiors.

In CCSNe, the MRI develops in the strongly differentially rotating layers around the newly formed PNS, where the condition $\partial_\varpi \Omega<0$ is generally satisfied in rapidly rotating progenitors \cite{Akiyama2003MRI}. Pre-collapse fields of ${\sim}10^{9}{-}10^{11}\,\mathrm{G}$, which are compressed during collapse, can then be exponentially amplified by the MRI and reach magnetar-class strengths of $10^{15}{-}10^{16}\,\mathrm{G}$ within a few milliseconds after bounce \cite{Obergaulinger2009Semiglobal,Mosta2015Dynamo,ReboulSalze2021GlobalMRI,Raynaud2020Magnetar}. The resulting MHD turbulence converts rotational energy into thermal and magnetic energy, and is thought to aid the stalled neutrino-driven shock \cite{Thompson2005Viscosity,Burrows2007MagneticallyDriven}. It may also power magnetorotational explosions associated with hypernovae (an exceptionally energetic type of stellar explosion) and the most energetic long gamma-ray bursts \cite{AloyObergaulinger2007,Bugli2020NonDipolar,Bugli2021Complex,AloyObergaulinger2021,Powell2023Magnetorotational}. 
However, a central numerical obstacle is that the characteristic wavelength of the dominant MRI mode, $\lambda_{\rm MRI}= 2\pi v_A{}^z/\Omega$, is of order meters for realistic seed fields. Here $v_A{}^z$ is the local Alfv\'en speed and $\Omega$ is the angular velocity, while $\lambda_{\rm MRI}$ is a proxy for (though not in general identical to) the true fastest-growing-mode wavelength. This is orders of magnitude below the grid scale of any feasible global simulation. As such, the quantitative results on growth, saturation, and channel-flow disruption (the parasitic destruction of the channel-flow structures that emerge during the linear MRI growth phase by secondary Kelvin-Helmholtz and tearing-mode instabilities) have been obtained primarily through local and semi-global simulation studies that resolve the linear instability and follow its nonlinear evolution into MHD turbulence \cite{Obergaulinger2009Semiglobal}.

In BNS mergers, the differentially rotating remnant typically exceeds the maximum mass supportable by any uniformly rotating configuration of the same equation of state \cite{ShibataUryu2000,Giacomazzo2011Collapse}. It is, therefore, expected to collapse as the differential rotation is suppressed by internal forces. During this process, three mechanisms amplify the magnetic field and redistribute angular momentum. First, the Kelvin-Helmholtz instability (KHI) acts at the shear interface immediately after merger and has been shown to amplify the field to magnetar-level strengths in high-resolution GRMHD simulations \cite{Kiuchi2015KelvinHelmholtz,Kiuchi2018Global,PalenzuelaEtAl2022,AguileraMiret2023Winding,AguileraMiret2025Robustness,Gutierrez2026Dynamo}. Magnetic winding then converts the remaining poloidal field into a stronger toroidal component on the Alfv\'en timescale \cite{Duez2006Evolution,Baiotti2008AccurateEvolutions,Shibata2021Winding}, while the MRI drives turbulent angular-momentum transport once the field becomes dynamically important \cite{Duez2006Evolution,Siegel2013,Kiuchi2023MassEjection}. MRI activity in hypermassive neutron star (HMNS) interiors was demonstrated by Siegel {et al.}, who found channel-flow structures with wavelengths and growth rates consistent with linear MRI theory \cite{Siegel2013}. Together, magnetic winding and MRI-driven transport can remove the support of the hypermassive remnant, leading to its collapse into a spinning black hole (BH) surrounded by a magnetized accretion torus \cite{Duez2006Evolution}, which is a leading central-engine model for SGRBs \cite{Eichler1989Nucleosynthesis,Narayan1992DeathThroes,Piran2004GRBPhysics,KumarZhang2015GRBJets,Sun2022Jet,Gottlieb2023Jets,CombiSiegel2023Jets,Bamber2024Jets,Musolino2025Neutrinos}.

A major difficulty in simulating the MRI in binary neutron star merger remnants is that its characteristic wavelength is extremely small. For realistic magnetic-field strengths of $B \sim 10^{12}\,\mathrm{G}$, the fastest-growing mode's wavelength is only $\lambda_{\rm MRI}{}^{\rm fgm} \sim \mathrm{cm}$, which is far shorter than the resolution achievable in current global simulations, even with modern high-performance computing resources. As such, MRI-driven turbulence is generally resolved only in local or semi-global setups \cite{Kiuchi2015KelvinHelmholtz,Kiuchi2018Global}.

Therefore, general-relativistic large-eddy simulation (GRLES) methods are used, in which the large-scale evolution is simulated explicitly while the effects of unresolved small-scale turbulence are approximated through a subgrid model \cite{Radice2017GRLES,Carrasco2020GradientSGS}. Quantitative modeling of these systems requires the measurement of effective transport coefficients that govern angular-momentum redistribution, field amplification, and eddy-scale closure, quantities that must be extracted directly from the simulations and are connected to the subgrid prescriptions used in large scale models. 

Historically, the net effect of MRI-driven turbulent stresses has been captured by a single phenomenological parameter, $\alpha_{\rm vis}$ \cite{ShakuraSunyaev1973}. However, it has been shown through higher-order closures and shearing-box simulations that turbulent magnetorotational stresses are not linearly proportional to the local shear and vanish identically for outwardly increasing angular-velocity profiles \cite{Pessah2008Stresses}. This is in direct contradiction to the assumptions underlying the $\alpha$ prescription. Penna et al. further showed that even in steady-state accretion onto a BH, $\alpha$ varies substantially with radius and depends on the dimensionless shear rate \cite{Penna2013VariableAlpha}. A quantitatively meaningful description of MRI-driven turbulence in compact-object remnants therefore demands transport-coefficient measurements extracted directly from resolved GRMHD simulations of the same configurations that subgrid models aim to represent. 

Full GRMHD employing gradient-type subgrid models and calibrated turbulence models, fit to high-resolution GRMHD simulations of Kiuchi and collaborators have enabled long-term post-merger studies of nucleosynthesis, ejecta composition, and electromagnetic counterparts \cite{Kiuchi2018Global,Radice2020Calibrated,RadiceBernuzzi2023,Hayashi2022BHNS}. The reliability of subgrid-based simulations depends on transport coefficients which must ultimately be calibrated using fully resolved MRI simulations under physically realistic conditions~\cite{RadiceHawke2024Turbulence}.

In this work we measure these coefficients directly from three-dimensional ideal-GRMHD simulations of isolated, differentially rotating neutron stars performed with \textsc{GR-Athena++} \cite{Cook2023GRAthena,daszuta_grathena_2021,daszuta2024grathena++_mag,Daszuta2026Neutrino}, which couples constrained-transport GRMHD to the Z4c formulation of the Einstein equations on a block-based AMR grid. 
We study two complementary classes of initial models. In the first, which we call the {non-collapsing} models, the star remains stable against gravitational collapse for the full simulated evolution and settles into a long-lived statistically steady state. In the second, the {collapsing} models, the star collapses to a black hole, either promptly through a dynamical instability or secularly once magnetic winding and the MRI have drained the differential-rotation support~\cite{Duez2006Evolution,Siegel2013}.
We extract $(\ell_{\rm mix},\alpha_{\rm DYN},\nu_T,\alpha_{\rm vis})$ from (azimuthal) angle averaged Maxwell stress using a Smagorinsky-type subgrid framework and a consistent estimate of the local MRI wavelength. Comparing these two classes allows us to examine whether MRI turbulence (where it develops) behaves consistently regardless of whether the star ultimately collapses. This comparison provides calibration data for the subgrid models of BNS merger remnants.

The rest of this article is organized as follows. Section \ref{sec:background} begins by presenting the ideal GRMHD equations governing the resolved scales and the relativistic angular-momentum flux they define (Section \ref{sec:ideal_grmhd}), and reviewing the MRI responsible for driving the turbulence (Section \ref{sec:mri}), before introducing the statistical description of turbulence, the averaging operations we use throughout, and the associated closure problem (Section \ref{sec:turbulence_closure}). We then coarse-grain the ideal GRMHD equations to identify the unresolved turbulent stresses (Section \ref{sec:effective}). These stresses are subsequently closed using a relativistic eddy-viscosity prescription (Section \ref{sec:nuT}) together with a mean-field dynamo model (Section \ref{sec:dynamo}). Section \ref{sec:models} describes our numerical setup and equilibrium models, and Section \ref{sec:results} presents our results, with our conclusions summarized in the final section. Throughout this work, we adopt geometrized units where $G=c=1$, and the metric signature $(-,+,+,+)$. Greek indices $\mu,\nu,\dots\in\{0,1,2,3\}$ denote spacetime components, while Latin indices $i,j,k,\dots\in\{1,2,3\}$ denote spatial components. We use cylindrical coordinates $(\varpi,\phi,z)$ adapted to the rotating star, with $\varpi$ the cylindrical radius, $\phi$ the azimuthal angle increasing counterclockwise about the rotation axis, and $z$ the height above the equatorial plane.

\section{Background}
\label{sec:background}
A hot, magnetized, differentially rotating NS and the turbulence that develops in its interior is the object of interest in this work. Such a star can form due to the collapse of a massive stellar core or as a remnant of a BNS merger. The differential rotation in this NS is a large reservoir of free energy. Its evolution, and whether the system collapses to a BH or not, depends on how efficiently this energy is used and redistributed \cite{Duez2006Evolution,Siegel2013,Giacomazzo2011Collapse}. One of the main mechanisms of this energy redistribution is the MRI, by transporting the angular momentum outward and driving the turbulent flow in the system~\cite{BalbusHawley1991,Balbus2003EnhancedTransport}.

\subsection{Ideal GRMHD on the resolved scales}
\label{sec:ideal_grmhd}
We refer to structures captured on the computational grid, i.e. those spanning at least a few grid spacings, as the resolved scales. At the resolved scales, the dynamics follow the ideal GRMHD equations on a dynamical spacetime. These equations couple baryon-number and stress-energy conservation to Maxwell's equations under the assumption of perfect conductivity \cite{Cook2023GRAthena}. The same framework underlies the current generation of magnetized neutron-star-merger simulations \cite{Kiuchi2024Review,RadiceHawke2024Turbulence,MizunoRezzolla2025}. Here we summarize the formulation implemented in \textsc{GR-Athena++}, which forms the basis of the analysis that follows. We recall only the elements needed for our diagnostics and refer the reader to standard texts for the full $3+1$ GRMHD derivations, e.g. \cite{Baumgarte:2010ndz,RezzollaZanotti2013}.

In the $3+1$ (ADM) decomposition, the line element is
\begin{equation}
    \mathrm{d}s^{2}=-\alpha^{2}\mathrm{d}t^{2}
    +\gamma_{ij}\!\left(\mathrm{d}x^{i}+\beta^{i}\mathrm{d}t\right)\!\left(\mathrm{d}x^{j}+\beta^{j}\mathrm{d}t\right),
    \label{eq:adm_metric}
\end{equation}
with lapse $\alpha$, shift $\beta^{i}$, spatial metric $\gamma_{ij}$, and $\sqrt{-g}=\alpha\sqrt{\gamma}$. The future-pointing unit normal $n_{\mu}=(-\alpha,0,0,0)$ defines the Eulerian observer, in whose frame the matter primitives (the rest-mass density $\rho$, pressure $P$, Eulerian velocity $v^i$, and magnetic field $B^i$) are defined. The fluid four-velocity decomposes as $u^{\mu}=W(n^{\mu}+v^{\mu})$, with $v^{\mu}n_{\mu}=0$ and Lorentz factor $W=(1-\gamma_{ij}v^{i}v^{j})^{-1/2}$. The three-velocity measured by the Eulerian observer is then
\begin{equation}
    v^{i}=\frac{u^{i}}{W}+\frac{\beta^{i}}{\alpha}.
    \label{eq:eulerian_velocity}
\end{equation}
The matter fields are derived from the decomposition of the stress-energy tensor $T_{ab}$,
\begin{equation}
    T_{ab}=E\ n_a n_b + 2S_{(a}n_{b)}+S_{ab}
\label{matter_field_decomp}
\end{equation}
with Eulerian energy density $E=T_{ab}\ n^an^b$, momentum $S_i=-T_{bc}n^b(g^c{}_i + n^cn_i)$, and spatial stress $S_{ij}=T_{cd} (g^c{}_i + n^c n_i) (g^d{}_j + n^d n_j)$~\cite{daszuta_grathena_2021}. The code evolves the conserved momentum $S^i$ to get the spatial projection of the fluid four-velocity ($V^i$), as one of its fundamental primitives ~\cite{daszuta_grathena_2021, daszuta2024grathena++_mag}. From this we form the
Eulerian three-velocity $v^{i} = V^{i}/W$ that enters our calculation.

With the local frame and fluid velocity fixed in this way, the star's differential rotation can be characterized by the coordinate angular velocity, $\Omega$, and the dimensionless shear parameter, $q$, defined as
\begin{equation}
    \Omega=\frac{u^{\phi}}{u^{t}}=\frac{\mathrm{d}\phi}{\mathrm{d}t},
    \qquad
    q=-\frac{\mathrm{d}\ln\Omega}{\mathrm{d}\ln\varpi},
    \label{eq:angular_velocity_phys}
\end{equation}
where $\varpi$ is the cylindrical radius. As such, $q>0$ for an outward-decreasing rotation ($\partial_\varpi \Omega<0$). For Newtonian power-law profiles, $q=3/2$ is Keplerian and $q=2$ gives constant specific angular momentum. The same values carry over to circular geodesic orbits in the Schwarzschild metric, if $\varpi$ is the areal radius \cite{BPT1972}.

The magnetic field threading this rotating fluid evolves under the perfect-conductivity (ideal-MHD) assumption stated at the start of this section, namely that the electric field vanishes in the frame comoving with the fluid. The homogeneous Maxwell equation then gives the no-monopole constraint $\partial_{i}\mathcal{B}^{i}=0$ for the densitized field $\mathcal{B}^{i}=\sqrt{\gamma}\,B^{i}$. As such, the field itself is transported by the induction equation
\begin{equation}
    \partial_{t}\mathcal{B}^{i}
    +\partial_{j}\!\left[\mathcal{B}^{i}\!\left(\alpha v^{j}-\beta^{j}\right)
                        -\mathcal{B}^{j}\!\left(\alpha v^{i}-\beta^{i}\right)\right]=0.
    \label{eq:gr_induction}
\end{equation}
The field moves with the fluid and is stretched by its motion. In a differentially rotating flow it is wound into a largely azimuthal configuration, whose magnetic tension couples fluid particles at neighboring radii and can mediate angular momentum transport.

The stress this winding builds up is expressed through the comoving magnetic four-vector $b^{\mu}={}^{\ast}\!F^{\mu\nu}u_{\nu}$ (with $b^{\mu}u_{\mu}=0$), whose invariant square is
\begin{equation}
    b^{2}=b^{\mu}b_{\mu}=\frac{B^{2}}{W^{2}}+\left(B^{i}v_{i}\right)^{2},
    \quad B^{2}=\gamma_{ij}B^{i}B^{j}.
    \label{eq:b_squared}
\end{equation}
The total stress-energy tensor, in Heaviside-Lorentz units (i.e. with the factor of $4\pi$ absorbed into the field), is then simply
\begin{equation}
    T^{\mu\nu}=\left(\rho h+b^{2}\right)u^{\mu}u^{\nu}+\left(P+\tfrac{1}{2}b^{2}\right)g^{\mu\nu}-b^{\mu}b^{\nu},
    \label{eq:stress_energy}
\end{equation}
with rest-mass density $\rho$, pressure $P$, and specific enthalpy $h=1+\tfrac{P \Gamma}{\rho(\Gamma-1)}$ with $\Gamma =2$ for this study \cite{Cook_2026Turbulence}. The quantity our diagnostics ultimately target is the radial flux of azimuthal angular momentum, which is the mixed component $T^{\varpi}{}_{\phi}$. As $\delta^{\varpi}{}_{\phi}=0$, the isotropic-pressure term drops, and we are left with
\begin{equation}
    T^{\varpi}{}_{\phi}=\left(\rho h+b^{2}\right)u^{\varpi}u_{\phi}-b^{\varpi}b_{\phi},
    \label{eq:angular_momentum_stress}
\end{equation}
whose magnetic part $\left(T^{\varpi}{}_{\phi}\right)_{\rm mag}=-b^{\varpi}b_{\phi}$ is the Maxwell stress. With the outward-positive convention, $b^{\varpi}b_{\phi}<0$ corresponds to a positive outward magnetic angular-momentum flux.

The mixed component is an angular-momentum flux in a precise sense, which is why it is the quantity our diagnostics target. To the extent that the spacetime is approximately axisymmetric (an approximation we make precise in Sect.~\ref{sec:effective}), $\xi^{\mu}{}_\phi=(\partial_{\phi})^{\mu}$ is an approximate Killing vector, and
\begin{equation}
    j^{\mu}= T^{\mu}{}_{\nu}\,\xi^{\nu}{}_\phi=T^{\mu}{}_{\phi}
    \label{eq:ang_mom_current}
\end{equation} 
is an approximately conserved angular-momentum current, with $\nabla_{\mu}j^{\mu}\approx0$. The source $T^{\mu\nu}\nabla_{\mu}\xi_{\nu}$ vanishes exactly only for an exact Killing symmetry \cite{Paschalidis2017,Baumgarte:2010ndz}, so non-axisymmetric structure in the remnant breaks strict conservation. The radial component of this current is precisely $T^{\varpi}{}_{\phi}$ from Eq.~\eqref{eq:angular_momentum_stress}. The same flux appears in the Shakura-Sunyaev $\alpha$-disc model, where turbulent transport is represented by an effective kinematic viscosity $\nu_{T}$ (defined in Sect.~\ref{sec:nuT})~\cite{ShakuraSunyaev1973,FrankKingRaine2002}. There it equals $\nu_{T}\,\rho\,(\mathrm{d}\Omega/\mathrm{d}\ln\varpi)$, so a positive $\nu_{T}$ drives outward transport down the angular-velocity gradient, with viscous heating $\propto\nu_{T}(\mathrm{d}\Omega/\mathrm{d}\ln\varpi)^{2}$ \cite{Balbus2003EnhancedTransport}. 

Our goal is to measure $T^{\varpi}{}_{\phi}$ and the coefficients that parametrize it ($\nu_{T}$, $\alpha_{\rm vis}$, $\ell_{\rm mix}$, $\lambda_{\rm MRI}$, and $\alpha_{\rm DYN}$), and to determine how they depend on density and on one another.

\subsection{The magnetorotational instability}
\label{sec:mri}
The Maxwell stress is generated and sustained by the MRI, and a simple picture captures its mechanism~\cite{BalbusHawley1991,Balbus2003EnhancedTransport}. Consider two fluid elements on neighboring circular orbits, joined by a weak field line that acts like a spring. When $\Omega$ decreases outwardly, the inner element runs ahead and stretches this spring-like line. Magnetic tension then torques the inner element backward and the outer element forward, which transports angular momentum outward. The inner element drops to a smaller radius and spins up (gets faster), while the outer element rises and slows down. Their separation grows, so even a small perturbation is amplified. This positive feedback is the MRI.

The instability requires $\mathrm{d}\Omega^{2}/\mathrm{d}\ln\varpi<0$ (i.e. $q>0$) and a sufficiently weak field. The field's role is to couple the elements through tension. In the weak-field regime, the growth rate is nearly independent of field strength, but the fastest-growing wavelength is not. The full linear dispersion analysis behind this picture is given in \cite{BalbusHawley1998}. In neutron-star remnants, the MRI is one of the leading mechanisms that amplify the field and redistribute angular momentum after merger, alongside the Kelvin-Helmholtz instability and magnetic winding \cite{Siegel2013,Kiuchi2024Review,RadiceHawke2024Turbulence}. We are interested in the nonlinear, saturated turbulence the instability sustains. Its characteristic scale, however, is set already during linear growth.

That scale is set by the relativistic Alfv\'en speed, $v_A$, which follows from the magnetic inertia in Eq.~\eqref{eq:stress_energy},
\begin{equation}
    v_{A}^{2}=\frac{b^{2}}{\rho h+b^{2}},
    \label{eq:rel_alfven_speed}
\end{equation}
and the fastest-growing mode (FGM) appears when magnetic tension balances the shear, $k\,v_A\sim\Omega$. This balance gives the characteristic wavelength
\begin{equation}
    \lambda_{\rm MRI}\sim\frac{2\pi\,v_{A}}{|\Omega|},
    \label{eq:lambda_mri}
\end{equation}
which is the order-of-magnitude scale quoted for NS remnants \cite{Radice2020Calibrated,DuezEtAl2005,Duez2006Evolution}. It is the scale on which the turbulence is mainly driven, and it sets the natural scale of the mixing length (Sect.~\ref{sec:transport_definitions}).

Whether a simulation can follow this process at all is a question of resolution. The grid captures the FGM only if it spans $\lambda_{\rm MRI}$. We measure this with the quality factor $Q_{\rm MRI}=\lambda_{\rm MRI}/\Delta x$~\cite{Noble2010MRI}. A value $Q_{\rm MRI}\gtrsim10$ is a common lower bound for the linear phase, while resolving the nonlinear, turbulent phase demands far greater coverage, as the cascade involves structures well below $\lambda_{\rm MRI}$ \cite{HawleyGuanKrolik2011,Hawley_2013,Rembiasz2016}. As $\lambda_{\rm MRI}$ scales with the field strength, realistic weak seed fields in dense matter give MRI wavelengths of only a few meters. These are far below the grid spacings (currently) achievable in global simulations. The instability and its turbulent cascade therefore remain unresolved in general, and if we want to capture their effects then we could use a subgrid model~\cite{Radice2017GRLES,Radice2020Calibrated,Siegel2013,MiravetTenes2026Subgrid}. During linear growth, the MRI produces quasi-axisymmetric channel modes which are organized, approximately axisymmetric streams of fluid and magnetic field with alternating velocity and field directions. Three-dimensional parasitic instabilities, of Kelvin-Helmholtz or tearing type, then disrupt these channels and drive the transition to statistically saturated turbulence \cite{Obergaulinger2009Semiglobal,Rembiasz2016}. We measure our diagnostics in that saturated phase. During the linear growth the field amplifies exponentially and no quantity is in equilibrium, so any extracted coefficient is transient and depends on when it is sampled. Saturation is the statistically steady state where energy injection balances dissipation, which is the only regime in which a transport coefficient carries a well-defined, time-averaged meaning.

\subsection{Turbulence and the closure problem}
\label{sec:turbulence_closure}
The saturated state is fully turbulent, and turbulence spans a continuous range of scales. Kinetic energy is injected at the integral scale $\ell_{0}$ and cascades to ever-smaller eddies, until viscosity dissipates it at the dissipation scale $\ell_{\rm d}$ \cite{SagautCambon2018,RadiceHawke2024Turbulence}. Kolmogorov scaling gives $\ell_{\rm d}\sim\ell_{0}\,\mathrm{Re}^{-3/4}$ where $\mathrm{Re}=\ell_{0}v_{\ell_{0}}/\nu$ is the Reynolds number, $v_{\ell_{0}}$ is the characteristic velocity of the largest eddies, and $\nu$ is the microscopic kinematic viscosity.

A direct simulation would therefore need $N_{\rm grid}\sim(\ell_{0}/\ell_{\rm d})^{3}\sim\mathrm{Re}^{9/4}$ cells, at a cost that scales roughly as $\mathrm{Re}^{3}$. The Reynolds numbers of NS interiors make this infeasible~\cite{RadiceHawke2024Turbulence,Kiuchi2024Review}. The practical alternative is to model the averaged effect of the unresolved scales on the resolved flow. This is the basis of the large-eddy-simulation (LES) and subgrid-scale (SGS) approaches now widely used for magnetized NS mergers~\cite{Radice2017GRLES,Carrasco2020GradientSGS,Vigano2020GRMHDLES,PalenzuelaEtAl2022,Aguilera_Miret_2022,Izquierdo2024LES,MiravetTenes2026Subgrid,Shibata2021Resistive}. It rests on a statistical decomposition, together with a closure for the correlations that the averaging introduces. Comprehensive treatments of the energy cascade, the closure problem, and large-eddy modeling are given in, e.g., \cite{Pope2000,SagautCambon2018}, and of astrophysical MHD in \cite{KatoFukue2020}.

Let $\langle\,\cdot\,\rangle$ denote a statistical average taken over $\phi$ at fixed $(t,\varpi,z)$. We use two versions of it, chosen by the nature of the field being averaged. Field quantities that are not mass densities are averaged with the proper-volume weight, which avoids a bias toward the dense core,
\begin{equation}
    \langle Q\rangle_{\rm vol}(\varpi,z)
    =\frac{\displaystyle\oint Q\,\sqrt{\gamma}\,\mathrm{d}\phi}
          {\displaystyle\oint \sqrt{\gamma}\,\mathrm{d}\phi}.
    \label{eq:vol_weighted_avg}
\end{equation}
Quantities that carry the rest mass in the conservation laws are instead averaged with the conserved rest-mass weight $\rho W\sqrt{\gamma}$. This is the relativistic mass-weighted (Favre) average,
\begin{equation}
    \langle Q\rangle^{\rho}_{\rm vol}(\varpi,z)
    =\frac{\displaystyle\oint Q\,\rho W\sqrt{\gamma}\,\mathrm{d}\phi}
          {\displaystyle\oint \rho W\sqrt{\gamma}\,\mathrm{d}\phi}.
    \label{eq:mass_weighted_avg}
\end{equation}

Any field splits as $Q=\langle Q\rangle+\delta Q$ with $\langle\delta Q\rangle=0$, which is the Reynolds decomposition. Averaging a quadratic term then leaves a fluctuation correlation that the mean fields do not determine,
\begin{equation}
    \langle Q_{1}Q_{2}\rangle=\langle Q_{1}\rangle\langle Q_{2}\rangle+\langle\delta Q_{1}\,\delta Q_{2}\rangle,
    \label{eq:quadratic_average}
\end{equation}
and this is what we call the closure problem. Each new correlation, if modeled by its own evolution equation, spawns higher-order correlations. The chain has to be closed at some point by expressing the unresolved correlations through the mean fields \cite{SagautCambon2018}.

For density quantities we use the Favre average $\langle Q\rangle^{\rho}_{\rm vol}$ of Eq.~\eqref{eq:mass_weighted_avg} with its fluctuation written as $\delta_f Q=Q-\langle Q\rangle^{\rho}_{\rm vol}$. 
This average keeps the continuity and momentum equations in conservative form, leaving a single residual turbulent stress, and it reduces to the plain azimuthal average $\langle Q\rangle$ only where density does not correlate with $Q$ \cite{SagautCambon2018,Balbus2003EnhancedTransport,Penna2013VariableAlpha,Radice2020Calibrated}.

Now, we look at how the residual stress relates to the mean flow. The simplest closure is the eddy-viscosity (Boussinesq) hypothesis. Here the turbulent stress is taken proportional to the mean rate of strain, through a turbulent viscosity $\nu_{T}$ \cite{RadiceHawke2024Turbulence,SagautCambon2018}. On dimensional grounds, a viscosity is a velocity times a length and as such,
\begin{equation}
    \nu_{T}\sim\ell_{\rm mix}\,v_{\rm turb},
    \label{eq:nu_dim}
\end{equation}
where $\ell_{\rm mix}$ is the mixing length \cite{prandtl1949developed} and $v_{\rm turb}$ is the eddy velocity. For MRI-driven turbulence we adopt $v_{\rm turb}=c_{s}$ (subsonic) which gives the closure $\nu_{T}=\ell_{\rm mix}\,c_{s}$ used below~\cite{Radice2020Calibrated}.

\subsection{Effective equations for the resolved scales}
\label{sec:effective}
We now apply the above averaging to the field equations themselves. Coarse-graining the ideal GRMHD equations for statistically axisymmetric flows can be implemented by taking the azimuthal average $\langle\cdot\rangle$ over $\phi$ at fixed $t$, $\varpi$, and $z$. For this average to commute with the covariant derivative, the geometry ($\alpha$, $\beta^{i}$, $\gamma_{ij}$, and hence the connection) must be $\phi$-independent. This is an {approximation}, valid only as long as non-axisymmetric structure in the remnant is subdominant. We adopt it throughout and treat the averaged spacetime as a slowly varying axisymmetric background \cite{Radice2020Calibrated}, so that
\begin{equation}
    \nabla_{\mu}\langle T^{\mu\nu}\rangle=0.
    \label{eq:averaged_conservation}
\end{equation}
From here on, $\langle\cdot\rangle_{\rm vol}$ and $\langle\cdot\rangle^{\rho}_{\rm vol}$ are the volume-weighted and mass-weighted (Favre) azimuthal averages of Eqs.~\eqref{eq:vol_weighted_avg} and \eqref{eq:mass_weighted_avg}.

The averaged stress-energy is not that of the mean flow. Because $T^{\mu\nu}$ is built from quadratic products of the dynamical fields, the average of a product is not the product of the averages (see, e.g., Eq.~\eqref{eq:quadratic_average}). Every quadratic term retains a fluctuation correlation. Take the Maxwell stress as an example, where $\langle b^{\varpi}b_{\phi}\rangle_{\rm vol}=\langle b^{\varpi}\rangle_{\rm vol}\langle b_{\phi}\rangle_{\rm vol} +\langle\delta b^{\varpi}\,\delta b_{\phi}\rangle_{\rm vol}$ and the second term is the turbulent Maxwell stress. 

As a result, the averaged stress-energy tensor is not the stress-energy tensor evaluated with the averaged fields, $\langle T^{\mu\nu}\rangle\neq T^{\mu\nu}\big[\langle\rho\rangle,\langle v^{i}\rangle,\langle B^{i}\rangle,\dots\big]$, and the fluctuation correlations are new unknowns. To make these unknowns precise, we first define the resolved state. We do not obtain it by averaging the primitive variables, because the average commutes with the conservative form of the equations but not with the nonlinear conservative-to-primitive inversion. Instead, we follow the general-relativistic large-eddy construction \cite{Radice2017GRLES,Radice2020Calibrated}. The {resolved primitives} $(\hat\rho,\hat v^{\,i},\hat P,\dots)$ are those recovered by inverting the standard conservative-to-primitive map on the {averaged} conserved variables $\langle D\rangle,\langle S_{i}\rangle,\langle\tau\rangle,\langle\mathcal{B}^{i}\rangle$. 
The conserved momentum density $S_{i}$ carries the full relativistic inertia $(\rho h+b^{2})W^{2}$ rather than the rest mass $\rho$. As a result, $\hat v^{\,i}$ differs from just the rest-mass average $\langle\rho v^{i}\rangle/\langle\rho\rangle$. Likewise, $\hat\rho$ and $\hat P$ differ from $\langle\rho\rangle$ and $\langle P\rangle$, because the recovery is nonlinear. We write $T_{\rm res}^{\mu\nu}$ for the stress-energy (see Eq.~\eqref{eq:stress_energy}) evaluated on these resolved primitives. The {residual stress} is then
\begin{equation}
    \tau^{\mu\nu}=\langle T^{\mu\nu}\rangle-T_{\rm res}^{\mu\nu},
    \label{eq:residual}
\end{equation}
built entirely from fluctuation correlations, and it vanishes for a strictly laminar, axisymmetric, steady flow. Equation~\eqref{eq:residual} is the azimuthal angle averaged analogue of the residual-stress definition of the calibrated large-eddy model \cite{Radice2020Calibrated} (their Sect.~2, Eqs.~(13)-(16)), in which the subgrid stress is the difference between the averaged total flux and the flux rebuilt from the averaged fields. We exploit this property directly. 
We emphasize that we do not evolve a subgrid model in this work. Our simulations resolve the turbulence, and measure the correlations entering Eq.~\eqref{eq:residual} directly from the resolved data. These measurements are then expressed in the form of the closure tensors employed by subgrid models (Sects.~\ref{sec:nuT} and \ref{sec:dynamo}), which is what allows us to extract the corresponding transport coefficients and compare them with the prescriptions used in large-eddy simulations. 
Substituting Eq.~\eqref{eq:residual} into Eq.~\eqref{eq:averaged_conservation} gives the effective equation $\nabla_{\mu}T_{\rm res}^{\mu\nu}=-\nabla_{\mu}\tau^{\mu\nu}$. The resolved flow obeys the ideal equations, but now {sourced} by the divergence of the residual stress. In practice we measure the residual transport as the azimuthal angle averaged Maxwell stress (Sect.~\ref{sec:transport_definitions}) and the fluctuation electromotive force discussed below.

For the angular-momentum evolution equation, the mixed residual stress splits term by term, through Eq.~\eqref{eq:quadratic_average}, into a relativistic Reynolds (kinetic) channel and a turbulent Maxwell channel,
\begin{equation}
    \tau^{\varpi}{}_{\phi} =
                            \underbrace{\big\langle\,(\rho h+b^{2})\,\delta_f(Wv^{\varpi})\,\delta_f(Wv_{\phi})\big\rangle_{\rm vol}}_{\mathrm{Reynolds}}
                            + \underbrace{\big\langle -b^{\varpi}\, b_{\phi}\big\rangle_{\rm vol}}_{\mathrm{Maxwell}}.
    \label{eq:residual_channels}
\end{equation}
Because $T_{\rm res}^{\mu\nu}$ retains the resolved (mean) velocity, the mean-flow kinetic stress is already captured there, so only the fluctuating part remains in the Reynolds channel.
In saturated MRI turbulence, the Maxwell stress is expected to dominate the Reynolds stress \cite{BalbusHawley1998,Balbus2003EnhancedTransport}. On this basis we neglect the kinetic Reynolds channel and build the transport diagnostics on the mixed Maxwell stress $-b^{\varpi}b_{\phi}$ alone. We keep the total Maxwell stress, both its mean and fluctuating parts. In our construction the resolved stress $T_{\rm res}^{\mu\nu}$ is evaluated with $b^{\mu}=0$, so the entire magnetic stress is carried by the residual (subgrid) stress. A subgrid model for a merger remnant must represent the effect of the full magnetic field, not only its turbulent part, because the fraction of the stress carried by the mean field varies from place to place \cite{Penna2013VariableAlpha}. We therefore do not subtract any mean-field contribution from $\langle -b^{\varpi}b_{\phi}\rangle_{\rm vol}$. We adopt this Maxwell-dominated approximation as a modeling assumption, not a measured property of these runs, consistent with the subsonic, subrelativistic turbulence assumed in calibrated subgrid models \cite{Radice2020Calibrated}.

The same coarse-graining of the induction equation~\eqref{eq:gr_induction} closes the magnetic evolution equation by producing a turbulent electromotive force (EMF). With $\mathcal{B}^{i}$ and the Eulerian velocity $v^{i}$ fluctuating, the mean field obeys
\begin{equation}
\begin{aligned}
    \partial_{t}\langle\mathcal{B}^{i}\rangle_{\rm vol}
    +\partial_{j} \Big[\langle\mathcal{B}^{i}& \rangle_{\rm vol}(\alpha \langle v^{j}\rangle_{\rm vol}-\beta^{j})\\
                       &-\langle\mathcal{B}^{j}\rangle_{\rm vol}(\alpha\langle v^{i}\rangle_{\rm vol}-\beta^{i})
                       +\mathcal{E}^{ij}\Big]
                        =0,
\end{aligned}
    \label{eq:mean_induction}
\end{equation}
where $\mathcal{E}^{ij}=\alpha\big(\langle\delta\mathcal{B}^{i}\delta v^{j}\rangle_{\rm vol} - \langle\delta\mathcal{B}^{j}\delta v^{i}\rangle_{\rm vol}\big)$. In three-vector form, this is the curl-producing correlation $\langle\bm{\mathcal{E}}\rangle_{\rm vol} = \langle\delta\bm{v}\times\delta\bm{B}\rangle_{\rm vol}$, with the fluctuation defined by $\delta Q = Q-\langle Q\rangle_{\rm vol}$. Unlike the residual stress, the EMF involves the {linear} (Reynolds) velocity fluctuations, because the field is not mass-weighted. Its standard closure is the mean-field $\alpha$-effect. To leading order, we expand the EMF in the mean field and its gradient,
\begin{equation}
    \langle\mathcal{E}_{i}\rangle_{\rm vol}=\alpha_{\rm DYN}\,\langle B_{i}\rangle_{\rm vol}
    -\eta_{\rm turb}\,(\nabla\times\langle B\rangle_{\rm vol})_{i}+\dots,
    \label{eq:alpha_DYN}
\end{equation}
where the pseudo-scalar $\alpha_{\rm DYN}$ (the $\alpha$-effect) measures the generation of large-scale field by helical small-scale turbulence, and $\eta_{\rm turb}$ is the turbulent diffusivity \cite{BrandenburgSubramanian2005,Moffatt1978,KrauseRadler1980}. In this context, helical means the eddies carry a preferred sense of twist, quantified by the kinetic helicity $\mathcal{H} = \langle \mathbf{v} \cdot (\nabla \times \mathbf{v})\rangle$, which in a rotating remnant is driven by the Coriolis force acting on the turbulence. Mean-field dynamos of this kind have recently been identified in BNS-merger remnants, where a turbulent $\alpha\Omega$ dynamo builds the large-scale field that ultimately drives outflows \cite{KiuchiEtAl2024,Most2023}. The vector $\langle\bm{\mathcal{E}}\rangle_{\rm vol}$ is polar while $\langle\bm{B}\rangle_{\rm vol}$ is axial, so $\alpha_{\rm DYN}$ is a pseudo-scalar. It is odd under reflection through the equatorial plane, so it changes sign across the mid-plane and vanishes on it. Equations~\eqref{eq:residual} and \eqref{eq:mean_induction} therefore introduce the two unclosed objects of this work, $\tau^{\mu\nu}$ and $\langle\bm{\mathcal{E}}\rangle_{\rm vol}$. We now close them with coefficients measured from the resolved flow.

\subsection{Transport coefficients and their measurement}
\label{sec:transport_definitions}
Rather than prescribe the residual stress and EMF, we measure them from each \textsc{GR-Athena++} time-snapshot. As such, we first fix the averaging operation. We then define each coefficient, together with the estimator used to evaluate it.

\paragraph{Center of mass and the two azimuthal averages.}
We interpolate all fields onto a uniform cylindrical grid $(\varpi,\phi,z)$, re-centred at every snapshot on the instantaneous baryonic-centre of mass,
\begin{equation}
    x^{i}_{\rm com}=\frac{\int x^{i}\,\rho W\sqrt{\gamma}\,\mathrm{d}^{3}x}{\int \rho W\sqrt{\gamma}\,\mathrm{d}^{3}x},
    \label{eq:com}
\end{equation}
with vectors and rank-2 tensors transformed by the corresponding Jacobian. The integrand $\rho W\sqrt{\gamma}=D\sqrt{\gamma}$ is the conserved proper rest-mass element, which is exactly the mass-weight of the Favre average. We then apply the two averages of Eqs.~\eqref{eq:vol_weighted_avg} and \eqref{eq:mass_weighted_avg}, choosing one or the other by the nature of each field. We use the mass-weighted average $\langle\cdot\rangle^{\rho}_{\rm vol}$ for $c_{s}^{2}$, $\Omega$, and $\lambda_{\rm MRI}$, and the volume-weighted average $\langle\cdot\rangle_{\rm vol}$ for the field quantities $\alpha_{\rm vis}$, $\nu_{T}$, $v_{A}{}^{z}$, and the dynamo means $\langle v^{i}\rangle_{\rm vol}$, $\langle B^{i}\rangle_{\rm vol}$, and $\langle\mathcal{E}_{i}\rangle_{\rm vol}$. Furthermore, the fluctuation of a field is taken as $\delta Q=Q-\langle Q\rangle_{\rm vol}$. Finally, for the thermodynamics, we use a $\Gamma$-law equation of state with $\Gamma=2$ and this gives $h = 1 + {2P}/{\rho}$ leading to 
\begin{equation}
    c_{s}^{2} =\frac{\Gamma P}{(\rho h)}=\frac{2P}{\rho+2P}.
\end{equation}

\subsubsection{Shakura-Sunyaev normalization \texorpdfstring{$\alpha_{\rm vis}$}{alpha\_vis}}
\label{sec:alpha_vis}
Normalizing the mean mixed Maxwell stress by the pressure gives the relativistic analogue of the classical
accretion-disc parameter \cite{ShakuraSunyaev1973},
\begin{equation}
    \alpha_{\rm vis}
    =\frac{\big\langle -\,b^{\varpi}b_{\phi}\big\rangle_{\rm vol}}{\big\langle P\big\rangle_{\rm vol}},
    \label{eq:alpha_vis}
\end{equation}
where both averages are volume-weighted. This ties the turbulent transport directly to the viscous-disc framework, in which $\alpha_{\rm vis}$ is the single free parameter \cite{Balbus2003EnhancedTransport,Penna2013VariableAlpha,FrankKingRaine2002}.

\subsubsection{Turbulent viscosity \texorpdfstring{$\nu_{T}$}{nu\_T} and mixing length
\texorpdfstring{$\ell_{\rm mix}$}{l\_mix}}
\label{sec:nuT}
We close the trace-free residual stress with the relativistic eddy-viscosity (Smagorinsky-type) model \cite{Radice2020Calibrated,Smagorinsky1963}. Let $D_{i}$ be the covariant derivative of $\gamma_{ij}$, with connection $\Gamma^{k}{}_{ij}=\tfrac12\gamma^{kl}(\partial_{i}\gamma_{jl}+\partial_{j}\gamma_{il} -\partial_{l}\gamma_{ij})$. The trace-free rate of strain of the mean Eulerian velocity is
\begin{equation}
    \Sigma_{ij}=\frac12\!\left(D_{i}v_{j}+D_{j}v_{i}\right)-\frac13\!\left(D_{k}v^{k}\right)\gamma_{ij}.
    \label{eq:rate_of_strain}
\end{equation}
We define the strain-based stress $\sigma_{ij}=-2(e+P)W^{2}\Sigma_{ij}$, with $e+P=\rho h$, and model the residual stress as $\tau^{\varpi}{}_{\phi}=\nu_{T}\,\sigma^{\,\varpi}{}_{\phi}$, with $\sigma^{\,\varpi}{}_{\phi}=\gamma^{\varpi k}\sigma_{k\phi}$. Now, equating $\tau^{\varpi}{}_{\phi}$ to the resolved Maxwell stress and inverting for the viscosity, we obtain
\begin{equation}
    \nu_{T}
        =\frac  {\big\langle -\,b^{\varpi}b_{\phi}\big\rangle_{\rm vol}}
                {\big\langle\, {\sigma}^{\,\varpi}{}_{\phi}\big\rangle_{\rm vol}},
    \quad
    \ell_{\rm mix}
        =\frac  {\nu_{T}}{\sqrt{\langle c_{s}^{2}\rangle^{\rho}_{\rm vol}}},
    \label{eq:nuT_lmix}
\end{equation}
where $\nu_{T}$ is a ratio of volume-weighted field averages and the sound speed is mass-weighted. This is the operational form of the closure $\nu_{T}=\ell_{\rm mix}c_{s}$. In the Newtonian limit ($W\to1$, $\rho h\to\rho$), it reduces to (Maxwell stress)$/(\rho\times\text{shear})$. For MRI-driven turbulence we expect $\ell_{\rm mix}$ to be of order the driving scale $\lambda_{\rm MRI}$ \cite{Radice2020Calibrated}. We evaluate this scale from the weak-field, vertical estimate 
\begin{equation}
    \lambda_{\rm MRI}=
        \frac{2\pi\,v_{A}{}^{z}}{|\Omega|},
    \quad 
    v_{A}{}^{z}=
        \frac{|B^{z}|}{\sqrt{4\pi\rho}},
    \label{eq:lambda_proxy}
\end{equation}
taken cell-by-cell and then mass-weighted averaged, following common practice for NS remnants \cite{Radice2020Calibrated,Siegel2013,Duez2006Evolution}. We use $\lambda_{\rm MRI}$ as a representative order-of-magnitude estimate of the FGM scale we can resolve, not as a single unstable wavelength. Here, $v_A{}^z$ is the Newtonian, weak-field limit of the relativistic Alfv\'en speed defined in Eq.~\eqref{eq:rel_alfven_speed}, evaluated in Gaussian units.

\subsubsection{Mean-field dynamo coefficient \texorpdfstring{$\alpha_{\rm DYN}$}{alpha\_DYN}}
\label{sec:dynamo}
We keep only the leading $\alpha$-term in Eq.~\eqref{eq:alpha_DYN} and project onto the mean field. This gives the least-squares estimate
\begin{equation}
    \alpha_{\rm DYN}
        =\frac  {\langle\mathcal{E}_{i}\rangle_{\rm vol}\,\langle B^{i}\rangle_{\rm vol}}
                {\langle B_{i}\rangle_{\rm vol}\,\langle B^{i}\rangle_{\rm vol}}.
    \label{eq:alpha_dyn_scalar}
\end{equation}
On the grid, we build the covariant turbulent EMF from the Eulerian velocity $v^{i}=V^{i}/W$ (Eq.~\eqref{eq:eulerian_velocity}) and the fluctuations $\delta Q=Q-\langle Q\rangle_{\rm vol}$. It is $\mathcal{E}_{i}=\epsilon_{ijk}\,\delta v^{j}\delta B^{k}$, with $\epsilon_{ijk}=\sqrt{\gamma}\,[ijk]$, so that
\begin{equation}
\begin{aligned}
    \mathcal{E}_{\varpi}&=\sqrt{\gamma}\left(\delta v^{\phi}\delta B^{z}-\delta v^{z}\delta B^{\phi}\right),\\
    \mathcal{E}_{\phi}  &=\sqrt{\gamma}\left(\delta v^{z}\delta B^{\varpi}-\delta v^{\varpi}\delta B^{z}\right),\\
    \mathcal{E}_{z}     &=\sqrt{\gamma}\left(\delta v^{\varpi}\delta B^{\phi}-\delta v^{\phi}\delta B^{\varpi}\right).
\end{aligned}
    \label{eq:emf_components}
\end{equation}
Volume-weighted averaging gives $\langle\mathcal{E}_{i}\rangle_{\rm vol}$ and $\langle B^{i}\rangle_{\rm vol}$, and Eq.~\eqref{eq:alpha_dyn_scalar} then gives the coefficient. Because $\alpha_{\rm DYN}$ is a pseudo-scalar that vanishes on the mid-plane, equatorial slices are physically uninformative for it. Its magnitude $|\alpha_{\rm DYN}|$ enters the density statistics on the same footing as the positive coefficients. A mean-field $\alpha\Omega$ dynamo of this kind has recently been measured in long-lived BNS-merger remnants \cite{KiuchiEtAl2024,Most2023,MostQuataert2023,CombiSiegel2023Jets,Kiuchi2026Magnetar}.

The coefficients $\nu_{T}$, $\ell_{\rm mix}$, $\alpha_{\rm vis}$, and $\alpha_{\rm DYN}$ parametrize the turbulence. We are essentially, measuring these coefficients, instead of using them in the evolution based on assumption as it is done in a subgrid model for two reasons. First, the eddy-viscosity closure is only approximately valid even in Newtonian turbulence \cite{RadiceHawke2024Turbulence,SagautCambon2018}. Second, the magnetorotational stress is not a true shear viscosity. It is not proportional to the local rate of strain, and it does not vanish where the strain does \cite{Pessah2008Stresses}. In addition, $\alpha_{\rm vis}$ is not universal and varies with radius and density, as in accretion discs \cite{Penna2013VariableAlpha}. 

\section{Models and numerical setup} \label{sec:models}

\subsection{Equilibrium configurations}
We evolve four differentially rotating neutron-star models, which we label \texttt{A1}, \texttt{A2}, \texttt{B1}, and \texttt{B2}. Models \texttt{A1}, \texttt{B1}, and \texttt{B2} follow the configurations of Duez {et al.}~\cite{Duez2006Evolution} (their stars A, B1, and B2), while \texttt{A2} is the sub-Kerr model A2 of Giacomazzo, Rezzolla \& Stergioulas~\cite{Giacomazzo2011Collapse}, subsequently evolved as a magnetized hypermassive neutron star by Siegel {et al.}~\cite{Siegel2013}. Our four simulations use a $\Gamma=2$ polytropic equation of state $P=K\,\rho^{\Gamma}$, with $K=100\ M_\odot^2$ and $\rho$ the rest-mass density. This sets the initial data only. During the evolution we adopt a $\Gamma$-law (ideal-gas) equation of state, $P=(\Gamma-1)\,\rho\,\epsilon$ with $\Gamma=2$, which matches the polytrope at $t=0$ but allows shock heating thereafter~\cite{Cook2023GRAthena,Giacomazzo2011Collapse}.

The rotation profile follows the standard ``$j$-constant'' law~\cite{Baumgarte:2010ndz},
\begin{equation}
  u^{t}\,u_{\phi} = A^{2}\,(\Omega_{c}-\Omega),  
  \label{eq:jconst}
\end{equation}
where $\Omega_{c}$ is the angular velocity on the rotation axis, $\Omega=u^{\phi}/u^{t}$, and $A$ is a constant length scale that sets the steepness of the differential rotation. In dimensionless form, $\hat{A}= A/\varpi_{e}$ with $\varpi_{e}$ the equatorial coordinate radius: $\hat{A}\to\infty$ recovers uniform rotation, while $\hat{A}\lesssim 1$ corresponds to strong differential rotation. We fix $\hat{A}=1$ (i.e.\ $A=\varpi_{e}$) throughout, as in Duez {et al.} The Newtonian limit of Eq.~\eqref{eq:jconst} is $\Omega=\Omega_{c}/(1+\varpi^{2}/A^{2})$, with $\varpi$ the cylindrical coordinate radius, so $\Omega$ decreases monotonically outward, the configuration the MRI requires.

The $j$-constant law is a mathematically convenient idealization rather than a faithful description of post-merger remnants. In general, better physically motivated rotation laws have since been derived and applied, calibrated against merger simulations, and our restriction to the $j$-constant family should be kept in mind when relating these models to realistic remnants~\cite{Hanauske2017,Uryu2017RotationLaws,Tsokaros2025Masking}.

The four models sample distinct rotational and stability regimes; we first fix the terminology. A {supramassive} star has a rest mass above the maximum non-rotating (TOV) value but can be supported by {uniform} rotation, up to the supramassive limit, $M_{\rm RNS}$, where a further increase in mass would require spin beyond the Keplerian limit---where the outer edge of the star becomes unbound due to rotation. A {hypermassive} star exceeds even this uniformly-rotating limit and is held up only by {differential} rotation~\cite{Baumgarte:2010ndz,Duez2006Evolution}. These mass classifications are independent of {dynamical stability}, i.e. stability against gravitational collapse on a dynamical (free-fall) timescale: a hypermassive star ($M_0>M_{\rm RNS}$) can still be dynamically stable, because its differential rotation supplies the extra support, and then collapse only {secularly}, once that support is removed. Dynamical stability is therefore {not} equivalent to $M_0<M_{\rm RNS}$.

Model \texttt{A1} (star A of~\cite{Duez2006Evolution}) is hypermassive: its rest mass exceeds the supramassive limit by $46\%$, with $J/M^{2}=1$~\cite{Duez2006Evolution}. It is dynamically stable and so does not collapse on a dynamical timescale; instead, magnetic winding and the MRI gradually redistribute its angular momentum and erode the differential-rotation support, driving a {secular} collapse to a BH. Its collapse is thus mediated by the very angular-momentum-transport processes this work quantifies. Model \texttt{A2} is, by contrast, hypermassive and dynamically unstable, sub-Kerr ($J/M^{2}<1$) and collapses promptly to a rotating BH. We adopt \texttt{A2} following Giacomazzo~et~al.~\cite{Giacomazzo2011Collapse} and Siegel~et~al.~\cite{Siegel2013}, as a comparison case with a different collapsing mechanism. Models \texttt{B1} and \texttt{B2} (stars B1 and B2 of~\cite{Duez2006Evolution}) are both non-hypermassive (rest mass below the supramassive limit) and dynamically stable. \texttt{B1} is {ultra-spinning}: its angular momentum exceeds the maximum attainable by a uniformly rotating star of the same rest mass before reaching the Keplerian limit, yet its rest mass stays below the uniform-rotation mass limit with a spin of $J/M^{2}=1$. As it is not hypermassive, magnetic angular-momentum transport does not drive it to collapse. Instead it redistributes angular momentum internally, settling into a new equilibrium with a nearly uniformly rotating core surrounded by a differentially rotating torus, in which the angular velocity is constant along magnetic field lines so that magnetic winding shuts down \cite{Duez2006Evolution}. Similar to \texttt{B1}, \texttt{B2} lies below the supramassive limit and does not collapse. However, with $J < J_{\max}$ ($J/M^{2} = 0.38$) it is not ultra-spinning. As a uniform rotation can exist at \texttt{B2}'s rest mass, the outward angular-momentum transport evens out its rotation and settles it into that uniform state unlike \texttt{B1}~\cite{Duez2006Evolution}. Therefore, \texttt{B2} undergoes the mildest evolution of our four models and serves as the baseline case. None of the four models is super-Kerr ($J/M^{2}>1$). Their initial properties are listed in Table~\ref{tab:model-properties}.

\begin{table}[tb!]
    \centering
    \setlength{\tabcolsep}{10pt}
    \begin{tabular}{@{}c|cccc@{}}
        \hline
        Quantity                  & \texttt{A1} & \texttt{A2} & \texttt{B1} & \texttt{B2} \\
        \hline
        $M$                       & 2.79        & 2.23        & 1.68        & 1.62        \\
        $M_0$                     & 3.04        & 2.39        & 1.78        & 1.76        \\
        $\rho_c\,(10^{-3})$       & 0.55        & 3.06        & 0.53        & 1.62        \\
        $r_p/r_e$                 & 0.31        & 0.33        & 0.46        & 0.85        \\
        $J$                       & 7.79        & 4.05        & 2.85        & 1.00        \\
        $J/M^2$                   & 1.00        & 0.82        & 1.00        & 0.38        \\
        $T/W$                     & 0.25        & 0.21        & 0.18        & 0.04        \\
        $\Omega_c\,(10^{-2})$     & 3.22        & 8.16        & 2.30        & 2.13        \\
        $\Omega_e\,(10^{-2})$     & 1.95        & 4.57        & 1.43        & 1.28        \\
        \hline
    \end{tabular}
    \caption{
        ADM mass $M$, baryon mass $M_0$, ADM angular momentum $J$, Central rest-mass density $\rho_c$, axis ratio $r_p/r_e$, dimensionless spin $J/M^2$, kinetic-to-binding energy ratio $T/W$, and central and equatorial angular velocities $\Omega_c$ and $\Omega_e$. Geometrized units with $G=c=1$ are used with mass expressed in $M_\odot$ which we follow in Table~\ref{tab:sim-parameters} as well.
        }
    \label{tab:model-properties}
\end{table}

\begin{table}[tb!]
    \centering
    \setlength{\tabcolsep}{8pt} 
    \begin{tabular}{c|cccc}
      \hline
      Model & \texttt{A1} & \texttt{A2} & \texttt{B1} & \texttt{B2} \\
      \hline
      EOS                         & $\Gamma=2$ & $\Gamma=2$ & $\Gamma=2$ & $\Gamma=2$ \\
      $R$                         & 12.59      & 5.28       & 13.61      & 7.84       \\
      $t_{\rm lim}$               & 8250       & 128        & 20700      & 7750       \\
      $L$                         & 204.8      & 64.0       & 204.8      & 102.4      \\
      $\Delta x_{\rm out}$        & 1.6        & 0.5        & 1.6        & 0.8        \\
      $l$                         & 18.89      & 7.92       & 20.42      & 11.76      \\
      $\Delta x$                  & 0.10       & 0.03125    & 0.10       & 0.05       \\
      $A_{\rm in}$                & 4.57       & 28.98      & 6.40       & 3.57       \\
      $P_{\rm cut}\,(10^{-5})$    & 0.21       & 6.44       & 0.11       & 1.04       \\
      $Q$                         & 6.52       & 100.33     & 10.96      & 39.75      \\
      \hline
    \end{tabular}
  \caption{All models use a $\Gamma=2$ polytropic EOS with $K=100$. $R$ is the stellar radius, and $t_{\rm lim}$ the final time. The outer domain $[-L,L]$ has $N=256$ points per direction with spacing $\Delta x_{\rm out}=2L/N$, and the finest grid $[-l,l]$ has spacing $\Delta x$. The initial field amplitude $A_{\rm in}$ and pressure cutoff $P_{\rm cut}$ are listed together with the quality factor $Q$ calculated at the point of maximum density at $t=0$.
  }
  \label{tab:sim-parameters}
\end{table}
All simulations use nested Cartesian static mesh-refinement grids with $16^{3}$ points per mesh-block. For each model the coarsest level covers $x^{i}\in[-L,L]$ with $N=256$ cells per spatial direction, giving an outer spacing $\Delta x_{\mathrm{out}}=2L/N$. Refinement is centered on the star; the finest level spans $x^{i}\in[-l,l]$ with spacing $\Delta x$. We choose $l=1.5R$ so that the stellar radius $R$ is well resolved, while the outer boundary $L$ is set by mesh structure to remain far enough from the star. This prevents reflections from contaminating the interior. Time integration uses a CFL factor $\Delta t/\Delta x=0.25$ on the finest grid, and each run is evolved to a final time $t_{\mathrm{lim}}$. The numerical parameters for all four runs are summarized in Table~\ref{tab:sim-parameters}.

Our models are initialized with a purely poloidal magnetic field, generated from the azimuthal component of the vector potential $A_{\phi} = \varpi^{2}\,\max\! \left[A_{\rm in}\,(P-P_{\mathrm{cut}}),\, 0\right]$ where $\varpi$ is again the cylindrical radius and the cutoff $P_{\mathrm{cut}}=0.04\times P_{\max}$ confines the field to the NS interior where $P_{\max}$ is the maximum pressure over the star. For hypermassive models \texttt{A1} and \texttt{A2}, the maximum density lies off the rotation axis, so $P_{\max}$ does not coincide with the central value $K\rho_c^{2}$. Furthermore, the normalization $A_{\rm in}$ sets the initial field strength, its value and the corresponding $P_{\mathrm{cut}}$ in code units are listed together with the grid parameters in Table~\ref{tab:sim-parameters}. 
    
\section{Results}
\label{sec:results}
We present the results in the order of the system's physical evolution. We start with the redistribution of angular momentum, which is driven by the magnetic-field amplification (Sect.~\ref{sec:res_evolution}). We will talk about the linear phase of the MRI and the channel modes and the spatial structure of the saturated turbulent state (Sect.~\ref{sec:res_channels_and_structure}). We then turn to the central result of this work: the time evolution of the transport coefficients (Sect.~\ref{sec:res_timeseries}), before examining how they depend on density and on one another (Sects.~\ref{sec:res_lmix_alphavis}, \ref{sec:res_alphadyn}, and \ref{sec:res_lmix_lambda}).

Unless stated otherwise, the saturated-phase quantities are time-averaged over the windows defined in Sect.~\ref{sec:res_timeseries} and indicated by the vertical dashed lines in the evolution figures (Fig.~\ref{fig:time_series} and Fig.~\ref{fig:time_series_appendix}).

\subsection{Magnetic-field amplification and angular-momentum redistribution}
\label{sec:res_evolution}
\begin{figure}[t]
  \centering
  \includegraphics[width=\columnwidth]{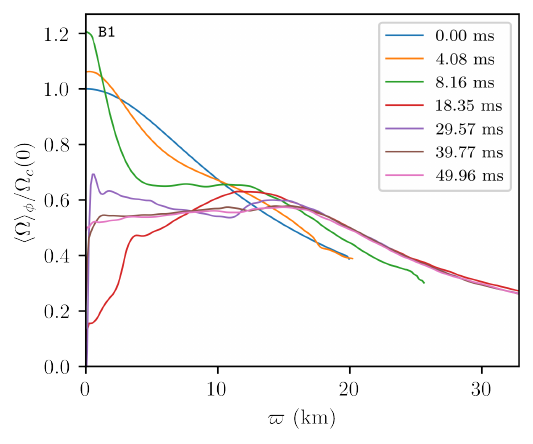}
  \caption{
  Angle (Azimuthal) averaged angular velocity profile for model B1, normalized by the initial central angular velocity $\Omega_c(0)$. The $\Omega$ profile is only shown where $\rho \geq 10^{-6}\,M_{\odot}^{-2}$. The profile decreases with the radius $\varpi$, indicating strong differential rotation. Initially we see that $\langle \Omega \rangle_\phi$ vanishes around $20\ {\rm km}$ as the star has not yet redistributed matter outward. However, magnetic stresses generated by winding and MRI-driven turbulence transport angular momentum from the rapidly rotating inner region to the more slowly rotating outer layers at larger radii. This flattens the angular-velocity gradient and we observe an increase in $\Omega$ in the bulk of the NS while also extending the profile outward.
  }
  \label{fig:omega_profile_b1}
\end{figure}

The stars are initialized with a strongly differential rotation profile, in which the angular velocity $\Omega$ decreases monotonically with cylindrical radius $\varpi$ (as defined in Sect.~\ref{sec:models}). This radial shear acts as the energy reservoir for the MRI amplification. Figure~\ref{fig:omega_profile_b1} shows the (azimuthal) angle-averaged profile $\Omega(\varpi)$ for model \texttt{B1}, normalized to the initial central value $\Omega_c(0)$. The initial profile is steeply peaked on the axis and falls off outward. Furthermore, at early times, it has no defined value beyond $\varpi \sim 20\ {\rm km}$ as we only average over cells with $\rho \geq 10^{-6}\,M_{\odot}^{-2}$, and at these early moments the star has not yet redistributed matter outward. As the evolution proceeds, magnetic winding and MRI-driven turbulence transport angular momentum outward to larger radii and the profile extends further out at later times. 

In Figure~\ref{fig:omega_profile_b1}, we see that, in the first few milliseconds, angular momentum moves away from the middle of the star ($\varpi\ \simeq 5 - 10\,{\rm km}$) towards both the innermost and outer reaches ($\varpi\gtrsim10\,\mathrm{km}$) of the star, as evidenced by an increase in angular velocity both at small and large radii. At $t=18.35\,{\rm ms}$, we see a sharp decrease in the angular velocity at the center of the star as angular momentum continues to be redistributed outward, resulting, after a further ${\sim}10\,\mathrm{ms}$, in a flatter $\Omega$ profile, indicating that, by this time, the rotation has become mostly rigid. The subsequent evolution indicates that the system has reached a quasi-steady state with a uniformly rotating inner-core.

\begin{figure*}
  \centering
    \includegraphics[width=\linewidth]{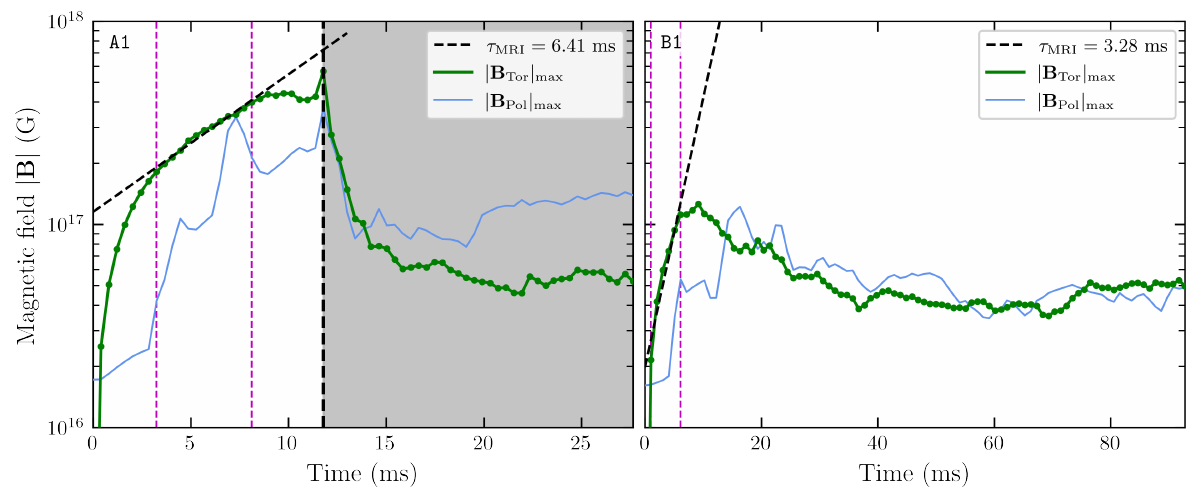}
  \caption{
  Evolution of the maximum toroidal and poloidal magnetic-field components (in log scale). The dashed black line shows the fitted exponential MRI growth with $\tau_\mathrm{MRI} = 6.41\,\mathrm{ms}$ (\texttt{A1}) and $3.28\,\mathrm{ms}$ (\texttt{B1}), over the MRI interval marked by the vertical dashed (magenta) lines. \texttt{B1} reaches a long-lived saturated state following amplification, while \texttt{A1}, despite a similar amplification interval, collapses shortly thereafter (shaded region).
  }
  \label{fig:B_evolution_all}
\end{figure*}

The magnetic amplification that drives this redistribution is shown in Fig.~\ref{fig:B_evolution_all}, which tracks the maximum toroidal and poloidal field components for \texttt{A1} and \texttt{B1}. We only show these models here as \texttt{B2} behaves similarly. All the models other than \texttt{A2} display the characteristic MRI fingerprint: an interval of exponential growth of the field, driven by the instability feeding off the shear, followed by saturation. \texttt{A2} collapses very early while at the start of MRI phase and does not develop any channel modes and associated turbulence characteristics. We fit the exponential to extract an $e$-folding time (the interval over which the field grows by a factor of $e$) $\tau_{\rm MRI} = 6.41\,\mathrm{ms}$ (\texttt{A1}),  $3.28\,\mathrm{ms}$ (\texttt{B1}) and $2.48\,\mathrm{ms}$ (\texttt{B2}) over the interval marked by the vertical dashed lines. Our growth times are broadly consistent with (but somewhat larger than) the full-GR axisymmetric simulations of Duez et al.~\cite{Duez2006Evolution}, who report $2.94\,\mathrm{ms}$ (\texttt{A1}) and $2.31\,\mathrm{ms}$ (\texttt{B1}) for the analogous models. These discrepancies may be attributable, at least in part, to the difference in dimensionality of the simulations. Three-dimensional, non-axisymmetric parasitic instabilities, absent in axisymmetric ($2D$) evolutions, disrupt the coherent channel growth earlier and may affect the extracted growth rate~\cite{Obergaulinger2009Semiglobal,Rembiasz2016}. A full analysis of the difference in both results, including other possible effects, is beyond the scope of this work.

After amplification, both \texttt{B1} and \texttt{B2} settle into a long-lived, statistically steady turbulent state, whereas \texttt{A1}, despite a comparable amplification phase, collapses to a BH shortly afterward (shaded region). This divergence in outcome, collapse for \texttt{A1} and \texttt{A2} versus a sustained turbulent state for \texttt{B1} and \texttt{B2}, is physical and follows directly from the models' stability classes (see discussion in Sect.~\ref{sec:models}).

\subsection{Channel modes, onset of turbulence and the spatial structure of the saturated state}
\label{sec:res_channels_and_structure}
\begin{figure*}
  \centering
    \includegraphics[width=\linewidth]{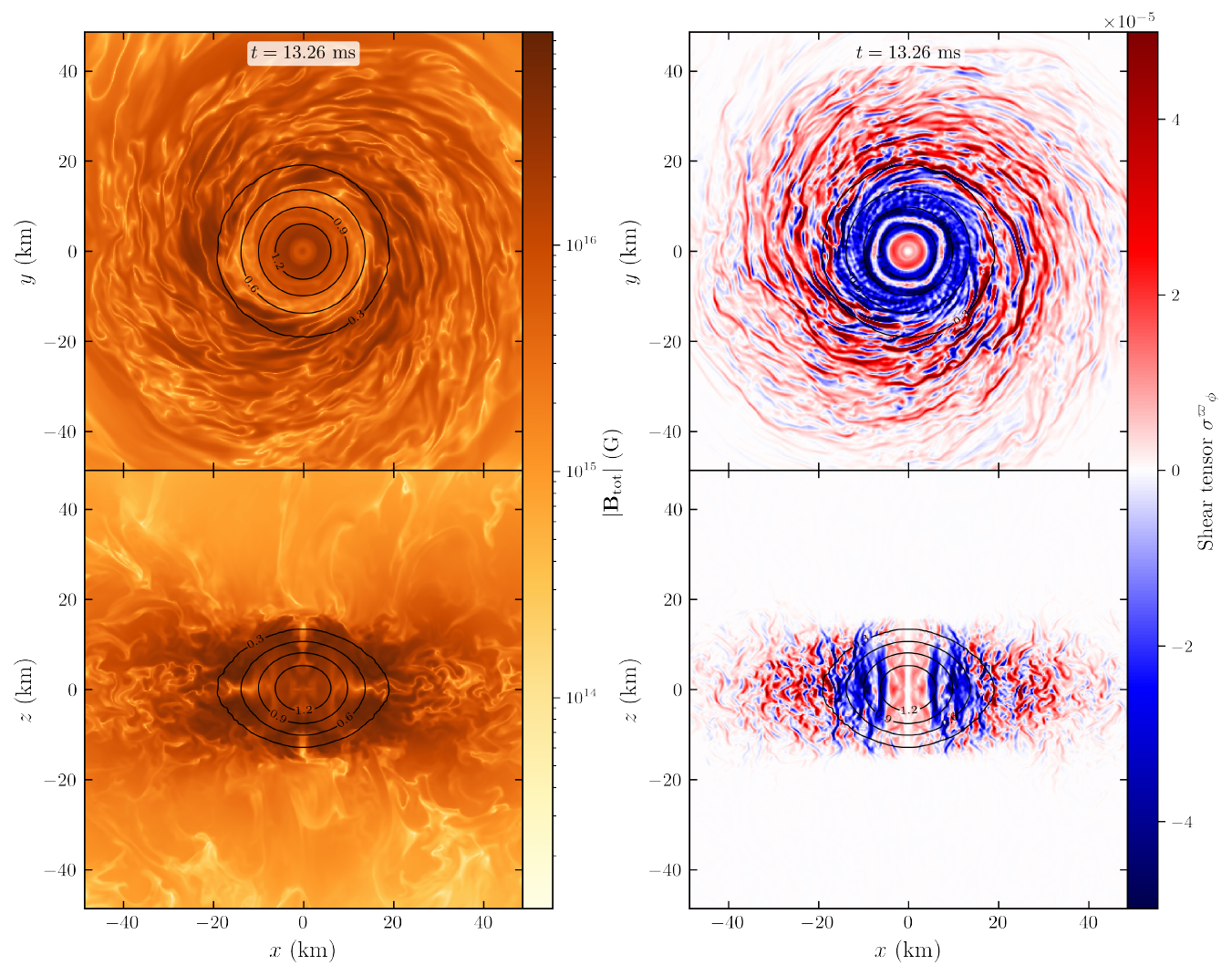}
  \caption{Total magnetic field $|\mathbf{B}|_{\rm tot}$ (left) and the mixed shear-tensor component $\sigma^{\varpi}{}_{\phi}$ (right) for \texttt{B1} at $t=13.26\,\mathrm{ms}$, shown in the equatorial ($xy$) and meridional ($xz$) planes. Contours show $\rho/\rho_c(0)$, with $\rho_c(0)$ the initial central density. The differential rotation winds $|\mathbf{B}|_{\rm tot}$ into a tight spiral in the equatorial plane, while $\sigma^{\varpi}{}_{\phi}$ shows radially alternating red and blue streams where neighboring layers shear in opposite directions. The well-defined, channel modes seen here are one of the expected signatures of resolved MRI although they need not coincide with the single fastest-growing mode.
  }
  \label{fig:channel_modes_B1}
\end{figure*}

During its linear phase, the MRI organizes the flow into {channel modes} as discussed in Sect.~\ref{sec:mri}. These are coherent, approximately axisymmetric streams of fluid and magnetic field with radially alternating velocity and field directions. Figure~\ref{fig:channel_modes_B1} captures this state for \texttt{B1} at $t = 13.26\,\mathrm{ms}$, just after the amplification phase, when the star is still contracting. In the equatorial ($xy$) plane the differential rotation winds the (total) magnetic field $\mathbf{B}_{\rm tot}$ into a tight spiral, while the meridional ($xz$) plane shows the field organized into alternating-sign layers stacked above and below the equator. The structure is clearest in the mixed shear-tensor component $\sigma^{\varpi}{}_{\phi}$, where radially adjacent red and blue streams shear in opposite directions, a direct kinematic imprint of the channels.

These channel modes are not turbulence, but its precursor. For each channel, neighboring layers move in opposite radial directions and stretch the field line connecting them, and the magnetic tension in that stretched field drives angular momentum outward. As the motion is coherent across the star, this transport is highly efficient. This same coherence, however, is what makes the channels fragile. The strong velocity and field gradients between neighboring streams are themselves unstable, and in three dimensions secondary parasitic instabilities of Kelvin-Helmholtz and tearing-mode type grow on these gradients and shred the channels apart
\cite{Obergaulinger2009Semiglobal,Rembiasz2016}. This breakdown converts the ordered, single-scale channel flow into a disordered cascade spanning a broad range of scales, which is the onset of MRI turbulence. The transport no longer proceeds through a few coherent streams but through a statistically steady field of eddies, and the angular-momentum flux fluctuates about a well-defined mean rather than growing without bound. It is this saturated turbulent state, not the transient channel phase, in which the transport coefficients acquire a stable, time-averaged meaning, and it is there that we measure them (Sect.~\ref{sec:res_timeseries}).

To model the MRI, we need to construct the grid with enough grid points for a chosen initial field strength such that the $\lambda_{\rm MRI}$ spans many cells. For our model \texttt{B1} the finest spacing $\Delta x = 147.7\,\mathrm{m}$ and the chosen seed field give a saturated MRI wavelength $\lambda_{\rm MRI}\approx 1.5\,\mathrm{km}$ giving $Q_{\rm MRI}\simeq 10$, which meets the benchmark commonly adopted for the linear phase, see Sect.~\ref{sec:mri}. Other models behave similarly (see Table~\ref{tab:sim-parameters}) except for \texttt{A1} which starts with $Q < 10$ but attains $Q\gtrsim10$ as the magnetic field gets amplified by magnetic winding early in the simulation. In Fig.~\ref{fig:channel_modes_B1}, the well-defined, multi-cell channel modes seen here are the expected signature of this resolved MRI, although the resolved scale need not coincide with the single fastest-growing mode. 

Once the parasitic instabilities disrupt the channels, the flow settles into fully developed, statistically steady turbulence. We illustrate the spatial structure of this saturated state with two transport coefficients for \texttt{B1} at $t = 55.06\,\mathrm{ms}$, well inside the averaging window of $37.73-80.55\ {\rm ms}$. 

\begin{figure}
  \centering
    \includegraphics[width=\linewidth]{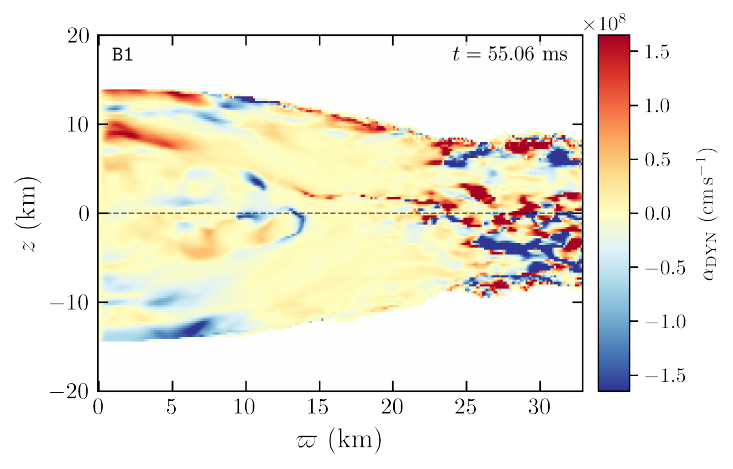}
    \caption{Dynamo coefficient $\alpha_{\rm DYN}$ for \texttt{B1} at $t=55.06\,\mathrm{ms}$ in the $r$-$z$ plane. The coefficient is positive (red) above the mid-plane and negative (blue) below, changing sign across the equator (dashed line, $z=0$). The $\alpha_{\rm DYN}$, changes sign across the equator and vanishes on it. This is phenomena proves the existence of a coherent $\alpha$-effect.}
  \label{fig:alpha_DYN_B1}
\end{figure}

Figure~\ref{fig:alpha_DYN_B1} shows the azimuthal angle averaged dynamo coefficient $\alpha_{\rm DYN}$, the component of the turbulent electromotive force aligned with the mean field (Eq.~\eqref{eq:alpha_dyn_scalar}). This field-aligned part is what drives the dynamo, because an EMF parallel to $\mathbf{B}$ sustains a current along the field that regenerates large-scale field. As $\alpha_{\rm DYN}$ is a pseudo-scalar, it is odd under reflection through the equatorial plane. The Coriolis force gives the turbulent eddies one sense of twist above the mid-plane and the opposite sense below. As such, the kinetic helicity, and with it the $\alpha_{\rm DYN}$, changes sign across the equator and vanishes on it. The clean anti-symmetry in Fig.~\ref{fig:alpha_DYN_B1} is therefore the mark of a coherent $\alpha$-effect rather than incoherent noise. In the dense interior the mean field is strong and the coefficient is smooth and large-scale, while in the low-density outer layers the mean field is weak and the projection defining $\alpha_{\rm DYN}$ breaks into small, sign-changing patches that are harder to measure cleanly.

\begin{figure}
  \centering
    \includegraphics[width=\linewidth]{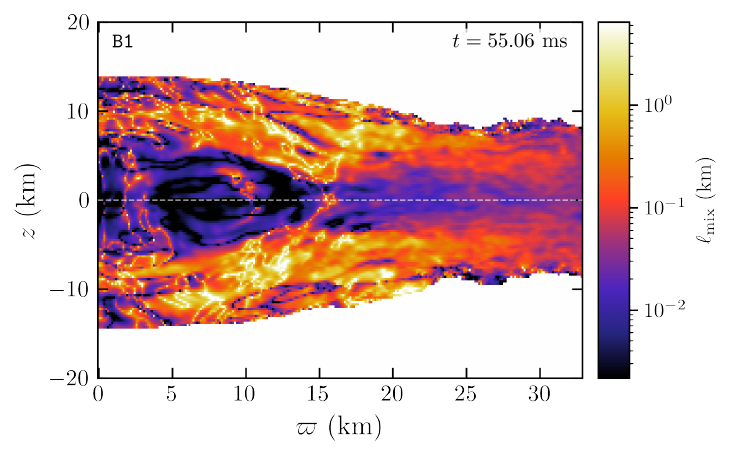}
    \caption{Turbulent mixing length $\ell_{\rm mix}$ for \texttt{B1} at $t = 55.06\,\mathrm{ms}$ in the $r$-$z$ plane. The fine, spiral-tangled filaments fill the bulk of the star and are the hallmark of fully developed turbulence.}
  \label{fig:lmix_B1}
\end{figure}

Similarly, Figure~\ref{fig:lmix_B1} shows the azimuthal angle averaged mixing length $\langle \ell_{\rm mix}\rangle_\phi^{\rm vol}$, the effective transport length defined through $\nu_T = \ell_{\rm mix}\,c_s$. The fine, spiral-tangled filaments filling the bulk of the star are the hallmark of fully developed turbulence. Both fields are spatially intricate but statistically stationary in this phase, which motivates the time-averaged, density-binned analysis that follows.

\subsection{Time evolution of the transport coefficients}
\label{sec:res_timeseries}
\begin{figure*}
  \centering
    \includegraphics[width=\linewidth]{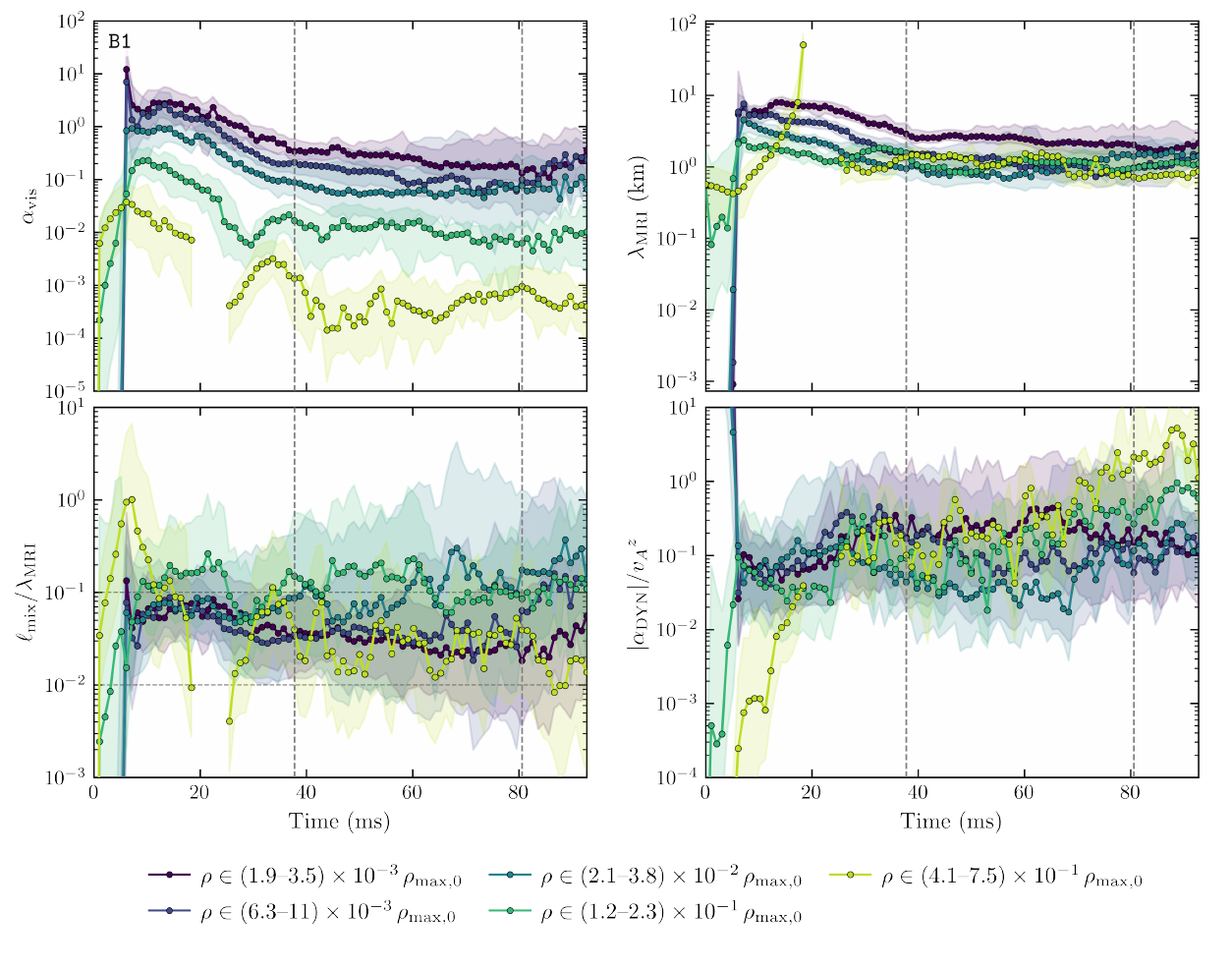}
  \caption{Time evolution of four transport coefficients for \texttt{B1}, with curves colored by density bin (in units of the initial maximum density $\rho_{\rm max,0}$). Clockwise from top left: the effective viscosity parameter $\alpha_{\rm vis}$, the MRI wavelength $\lambda_{\rm MRI}$, the dynamo measure $|\alpha_{\rm DYN}|/v_A{}^z$, and the ratio $\ell_{\rm mix}/\lambda_{\rm MRI}$. Vertical dashed lines mark the saturated window over which the coefficients are averaged. All of these quantities show a similar pattern: a sharp transient as the MRI grows and saturates within the first $\sim\!15\,\mathrm{ms}$, followed by a long, statistically steady turbulent state.}
  \label{fig:time_series}
\end{figure*}

Figure~\ref{fig:time_series} shows evolution of these coefficients for \texttt{B1}, while the other models are shown in Fig.~\ref{fig:time_series_appendix} in the Appendix~\ref{appendix}. Each curve corresponds to one density bin, colored by density in units of the initial maximum density $\rho_{\rm max,0}$. As such, the spread of colors at a given time shows how a coefficient varies across the star's density structure. The bins are ten globally fixed logarithmic intervals spanning $[\rho_{\min}^{\rm glob},\rho_{\max}^{\rm glob}]$, where $\rho_{\max}^{\rm glob}=0.75\,\rho_{\max,0}$  ($\rho_{\rm max,0}$ is the maximum density from the initial data) and $\rho_{\rm min}^{\rm glob} = 10^{-6} M_\odot^{-2}$ frozen from the initial-data snapshot. At each snapshot, the plotted curve is the rest-mass-density-weighted spatial median within the bin, and the shaded band encloses the region between the $16^{\rm th}$ and $84^{\rm th}$ percentiles. Bins with too few cells to resolve the percentiles are left empty. For the density profiles shown later in Sect.~\ref{sec:res_lmix_alphavis}, we time-average each bin over its model's saturated window (the vertical dashed lines). The positive-definite coefficients span orders of magnitude, so we aggregate in log space. The central value is $10^{\mu_{\log}}$ with $\mu_{\log}=N_{w}^{-1}\sum_{t}\log_{10}\widetilde{X}(\rho,t)$, the geometric mean of the per-snapshot medians over the $N_w$ snapshots in the averaging window, and the whiskers mark $10^{\mu_{\log}\pm\sigma_{\log}}$. Since $\alpha_{\rm DYN}$ is signed and antisymmetric about the mid-plane, $|\alpha_{\rm DYN}|$ enters this log-space statistic on the same footing as the other coefficients. All four quantities share the same broad temporal pattern with a sharp transient during the linear MRI growth, saturation within the first ${\sim}15\,\mathrm{ms}$ (\texttt{B1}), and then a long, statistically steady turbulent plateau over which we measure the coefficients (vertical dashed lines). We discuss each panel in turn.

The effective viscosity parameter $\alpha_{\rm vis}$ (top left) is the mean Maxwell stress normalized by pressure (Eq.~\eqref{eq:alpha_vis}). During the linear MRI growth it spikes sharply as the field amplifies. Once the channels break down and the turbulence saturates, it relaxes to a quasi-steady plateau that persists for the rest of the run. In this saturated state the curves are strongly ordered by density. The low-density bins settle near $\alpha_{\rm vis}\sim 10^{-1}$, while the densest bins sit several orders of magnitude lower. This vertical spread holds throughout saturation and is the first indication that the viscous stress carries a real density dependence rather than a single universal value. A non-universal, position-dependent $\alpha$ has likewise been reported for accretion discs, where it varies with radius and local shear \cite{Penna2013VariableAlpha}. We note further that although \texttt{B2} (Fig.~\ref{fig:time_series_appendix}) follows the same trend, its bins saturate at different values, so the saturated $\alpha_{\rm vis}$ from this closure also depends on the model itself.

The MRI wavelength $\lambda_{\rm MRI}$ (top right) (Eq.~\eqref{eq:lambda_proxy}) grows together with the field, because $\lambda_{\rm MRI}\propto v_A\propto |\mathbf{B}|$. As such, by the end of the MRI phase, it saturates at a roughly common value of order ${\sim}1\,\mathrm{km}$ (a few times the simulation grid spacing) across all density bins, after which it changes little. This saturation is expected on physical grounds. The MRI is a {local} instability, so it ceases to grow once $\lambda_{\rm MRI}$ becomes comparable to the size of the star, the field having amplified until the unstable mode no longer fits within it. In our runs that saturation scale is also comparable to only a few cells of the finest grid. The macroscopic value of $\lambda_{\rm MRI}$ should therefore be read as the characteristic scale the simulation resolves, not as the true, much smaller MRI wavelength of a realistic weak-field neutron star (Sect.~\ref{sec:mri}).

The ratio of the mixing length to the MRI wavelength $\ell_{\rm mix}/\lambda_{\rm MRI}$ (bottom left)  settles onto a narrow band after the transient. We observe that $\ell_{\rm mix}/\lambda_{\rm MRI}\sim 0.01-0.1$, which is both flat in time and nearly independent of density. This is a quantitatively useful result. In general the common subgrid assumption that the mixing length is of order the MRI driving scale, $\ell_{\rm mix}\sim\lambda_{\rm MRI}$~\cite{Radice2020Calibrated}, overestimates the transport length by roughly one to two orders of magnitude based on our saturated turbulence data.

The dynamo measure $|\alpha_{\rm DYN}|/v_A{}^z$ (bottom right) is the dimensionless ratio of the dynamo coefficient (which has units of velocity) to the vertical Alfv\'en speed. It gauges how vigorous the turbulent dynamo is relative to the Alfv\'enic speed. It rises during growth and then fluctuates around values of order $10^{-2}-1$ in saturation, with the density ranges intermixed and no clear ordering, indicating the absence of a clean density dependence for the dynamo $\alpha$ (Sect.~\ref{sec:res_alphadyn}). Figure~\ref{fig:time_series_appendix} confirms that this behavior is generic rather than specific to \texttt{B1}. Model \texttt{B2} is non-collapsing and reaches a sustained, quasi-steady turbulent state, reproducing the \texttt{B1} hierarchy of the four coefficients. Model \texttt{A1} develops the same plateau as the MRI phase ends and channel modes are formed. \texttt{A1}'s agreement during the start of the turbulent phase reinforces the result despite its shorter lifetime. By contrast, \texttt{A2} forms a black hole within $\lesssim 0.5\,\mathrm{ms}$, before the MRI can develop, so it never establishes a turbulent state. Siegel~et~al.~\cite{Siegel2013}, by comparison, report the formation of channel modes during the evolution of this model prior to its collapse. These features are not observed in our simulations of this model. A detailed study of the origins of this discrepancy are beyond the scope of the present work, however we note that there are fundamental differences in the evolution methods used in that and this work, e.g.~the `Constrained Transport' \cite{Evans:1988qd, White:2015omx} magnetic field evolution in \textsc{GR-Athena++} vs.~the so-called `Flux-CT'~\cite{Balsara1999} method of \textsc{WhiskyMHD}~\cite{Giacomazzo:2007ti}. The coefficients measured in \texttt{A2} reflect this absence of turbulence and show no saturated structure. Taken together, the saturated transport state is recovered in every model that sustains MRI turbulence and is absent only where collapse preempts it.

\subsection{The mixing length and the effective viscosity}
\label{sec:res_lmix_alphavis}
We now isolate how the mixing length $\ell_{\rm mix}$ and the effective viscosity $\alpha_{\rm vis}$ depend on rest-mass density $\rho$ and on each other, while also comparing an early channel-mode phase to the saturated turbulent phase.

\begin{figure*}
  \centering
    \includegraphics[width=\linewidth]{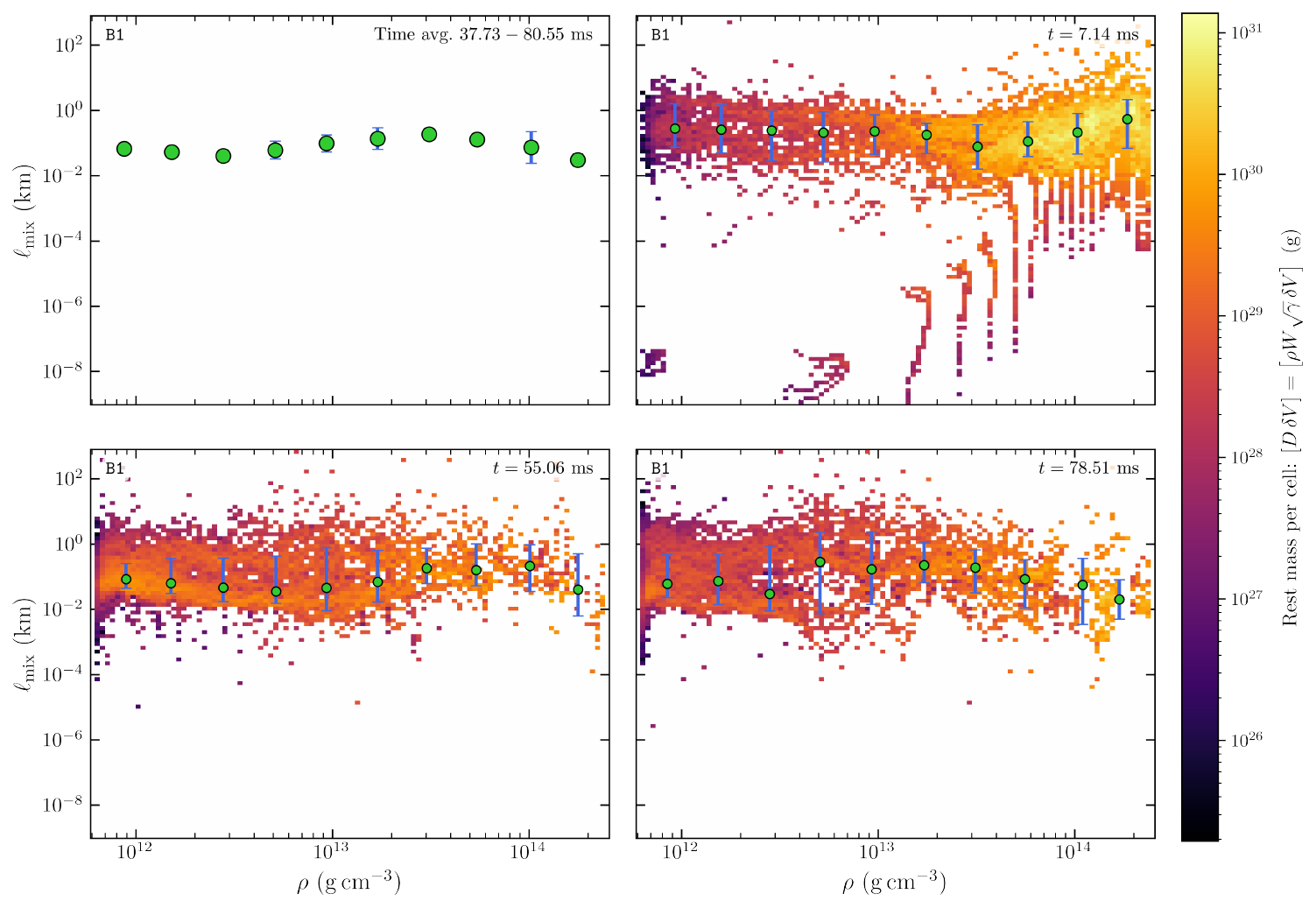}
    \caption{Mixing length $\ell_{\rm mix}$ versus rest-mass density $\rho$ for \texttt{B1}. Top left: time-averaged over the saturated window ($37.73-80.55\,\mathrm{ms}$). The remaining panels show the mass-weighted two-dimensional distribution with per-bin central values (points) at $t=7.14\,\mathrm{ms}$, $t=55.06\,\mathrm{ms}$, and $t=78.51\,\mathrm{ms}$. $\ell_{\rm mix}$ governed by the turbulent dynamics and is independent of the local density directly. $\ell_{\rm mix}$ is nearly constant across the star, holding a fixed fraction of the driving scale once the star reaches its steady state.}
  \label{fig:lmix_vs_rho}
\end{figure*}

Starting with the mixing length in Fig.~\ref{fig:lmix_vs_rho}, we see that the time-averaged profile is nearly flat across the full density range. The three snapshots show that during MRI growth ($t = 7.14\,\mathrm{ms}$) the distribution is still organizing and retains a mild upward trend at higher densities, but by the $t = 55.06\,\mathrm{ms}$ and $t = 78.51\,\mathrm{ms}$ snapshots it has settled into a broad band that is essentially independent of density and stable in time. Like $\alpha_{\rm vis}$, $\ell_{\rm mix}$ is set by the turbulent dynamics rather than by the local density directly. The difference is in what that dependence looks like once the turbulence saturates. Where $\alpha_{\rm vis}$ inherits a clean but model-dependent density trend, $\ell_{\rm mix}$ comes out nearly constant across the star, holding a fixed fraction of the driving scale, i.e.~$\ell_{\rm mix}\approx (10^{-2}-10^{-1})\,\lambda_{\rm MRI}$. This near-constancy is useful for subgrid modeling. Taking $\ell_{\rm mix}$ as a single representative value is a reasonable approximation, whereas $\alpha_{\rm vis}$ carries a density dependence that is not predictable in advance and likely reflects the details of the initial configuration.

\begin{figure*}
  \centering
    \includegraphics[width=\linewidth]{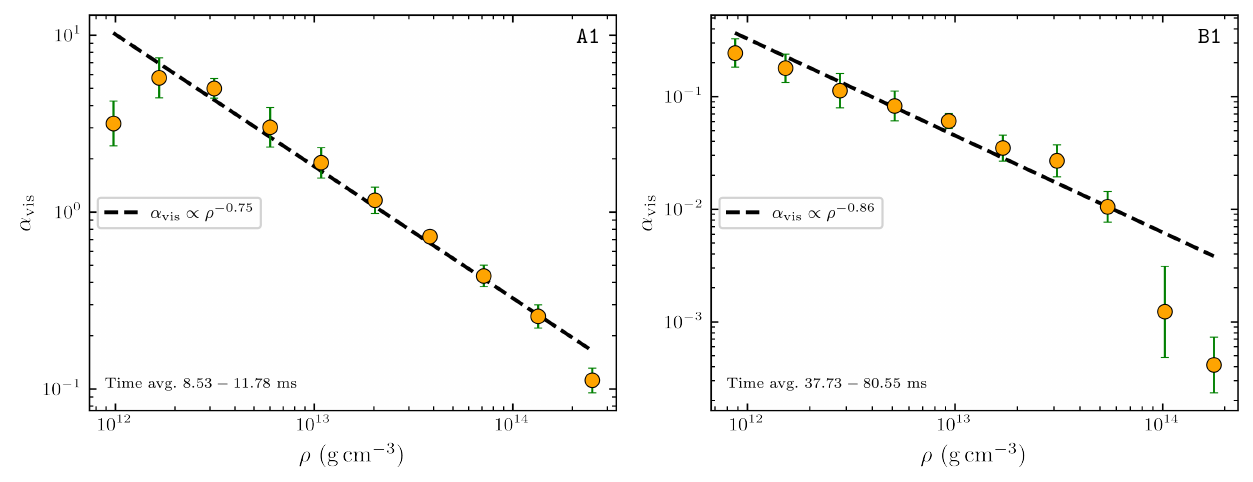}
  \caption{Effective viscosity parameter $\alpha_{\rm vis}$ versus rest-mass density $\rho$ for \texttt{A1} (left, averaged over $8.53$–$11.78\,\mathrm{ms}$) and \texttt{B1} (right, averaged over $37.73$–$80.55\,\mathrm{ms}$). Dashed black lines are power-law fits weighted by the per-bin uncertainties, with slopes $\alpha_{\rm vis}\propto\rho^{-0.75}$ (\texttt{A1}) and $\rho^{-0.86}$ (\texttt{B1}). The slope changes from model to model, so although the decline is robust, its exponent is not universal.
  }
  \label{fig:alpha_vis_vs_rho}
\end{figure*}

The effective viscosity $\alpha_{\rm vis}$ (see Fig.~\ref{fig:alpha_vis_vs_rho}) follows a clear, monotonic power-law {decline} with density that persists from the early phase into saturation, where $\alpha_{\rm vis}\propto\rho^{-0.75\,(-0.86)}$ for model \texttt{A1} (\texttt{B1}). However, \texttt{A1}'s slope is measured during its early channel-mode phase rather than a saturated state, as such, this coefficient should be read as transient rather than as a steady-state value. The dense, slowly evolving core thus sustains proportionally weaker effective viscosity than the lower density outer material. The appendix figure~\ref{fig:alpha_vis_vs_rho_A2_B2_appendix} extends this to the remaining models. \texttt{B2} reproduces the same monotonic decline with $\alpha_{\rm vis}\propto\rho^{-0.41}$, supporting the generality of the trend, while \texttt{A2}, which collapses before turbulence develops, never establishes a meaningful $\alpha_{\rm vis}$ and yields only a positive fitted slope. The slope therefore changes from model to model, so although the decline is robust, its exponent is not universal. This is consistent with accretion-disc simulations, where the Shakura-Sunyaev $\alpha$ is found to be a systematically varying, non-constant quantity rather than a fixed coefficient~\cite{Penna2013VariableAlpha,Shibata2017ViscousHydrodynamics}. We note that in that context the variation is reported with disc radius and local shear, which here maps onto our density coordinate. This is precisely why we measure $\alpha_{\rm vis}(\rho)$ rather than adopt a single value (see Sect.~\ref{sec:alpha_vis})~\cite{Pessah2008Stresses}.

\begin{figure}
\centering
    \includegraphics[width=\linewidth]{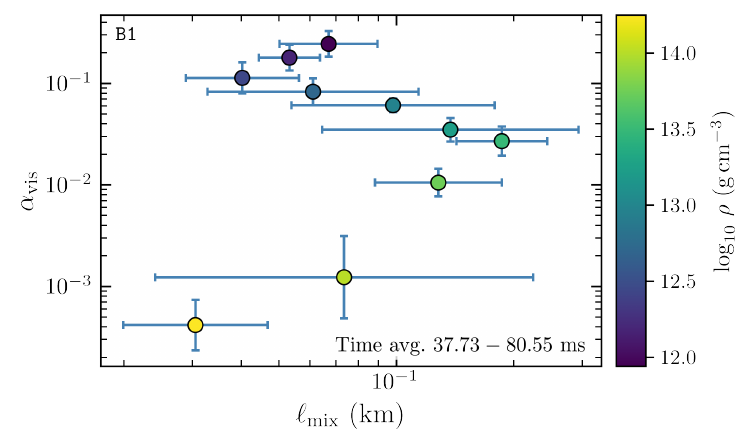}
    \caption{Mixing length $\ell_{\rm mix}$ against the effective viscosity parameter $\alpha_{\rm vis}$ for \texttt{B1}, time-averaged over the saturated window ($37.73$–$80.55\,\mathrm{ms}$), with points colored by density bin. From spread of points, we can see that there is no corelation between $\alpha_{\rm vis}$ and $\ell_{\rm mix}$.}
    \label{fig:lmix_vs_alpha_vis}
\end{figure}

Having established this density trend, we can ask whether $\alpha_{\rm vis}$ also correlates with the mixing length across the density bins, over the same saturated window. Figure~\ref{fig:lmix_vs_alpha_vis} plots $\alpha_{\rm vis}$ against $\ell_{\rm mix}$, and the two show no strong correlation. The mixing length is confined to a narrow saturated range while $\alpha_{\rm vis}$ spans roughly four orders of magnitude, and that spread is set by density, not by $\ell_{\rm mix}$, consistent with $\ell_{\rm mix}$ having converged to a fixed turbulent scale. We find the same in \texttt{B2}. As $\ell_{\rm mix}$ and $\alpha_{\rm vis}$ are both built from the same Maxwell stress, the absence of a relation between them is a non-trivial property of the saturated flow, not a statement about two unrelated quantities.

\subsection{The dynamo coefficient}
\label{sec:res_alphadyn}
\begin{figure*}
  \centering
    \includegraphics[width=\linewidth]{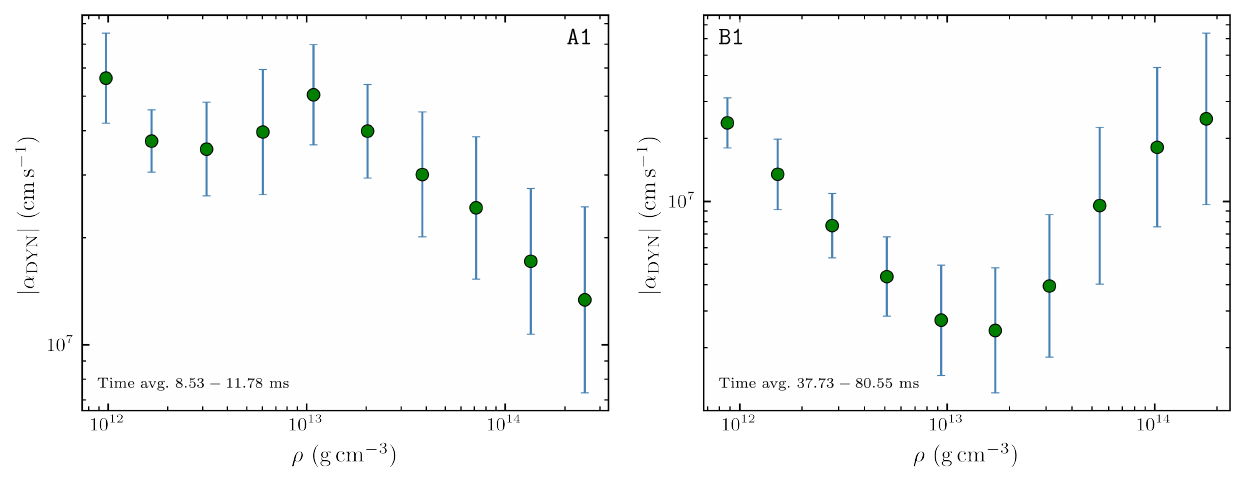}  
  \caption{$|\alpha_{\rm DYN}|$ as a function of rest-mass density $\rho$ for \texttt{A1} (left), time-averaged over the early channel mode phase ($8.53-11.78\,\mathrm{ms}$), and \texttt{B1} (right), time-averaged over the saturated turbulent phase ($37.73-80.55\,\mathrm{ms}$). Points are the per-bin central values and error bars their scatter over each window. \texttt{A1} shows a slight, near-monotonic decline in $|\alpha_{\rm DYN}|$ with density. However, \texttt{B1} is non-monotonic, with a minimum near $\rho\sim10^{13}\,\mathrm{g\,cm^{-3}}$ and an upturn toward both the lowest and highest densities. We find comparable behavior in model \texttt{B2}, indicating that $|\alpha_{\rm DYN}|$ retains no clean density dependence across models in the developed turbulent state.}
  \label{fig:alpha_dyn_vs_rho}
\end{figure*}

Figure~\ref{fig:alpha_dyn_vs_rho} shows $|\alpha_{\rm DYN}|$ against $\rho$. In every turbulent model the profile is non-monotonic, with a minimum near $\rho\sim 10^{13}\,\mathrm{g\,cm^{-3}}$ and an upturn toward both the lowest and highest densities. The shape varies between \texttt{A1}, \texttt{B1}, and \texttt{B2}, so $|\alpha_{\rm DYN}|$ carries no universal density dependence in the saturated turbulent state. The outermost, low-density bins are also where the mean field $\langle\mathbf{B}\rangle_\phi$ is weakest and the projection defining $\alpha_{\rm DYN}$ (Eq.~\eqref{eq:alpha_dyn_scalar}) is most susceptible to noise (Fig.~\ref{fig:alpha_DYN_B1}), so the upturn at the low-density end should be read with that in mind.

\begin{figure}
\centering
\includegraphics[width=\linewidth]{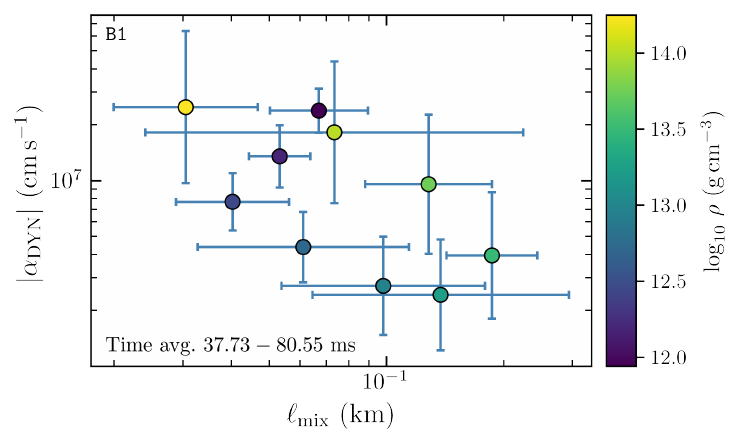}
    \caption{The dynamo coefficient $|\alpha_{\rm DYN}|$ against $\ell_{\rm mix}$ for \texttt{B1}, time-averaged over the saturated window ($37.73-80.55\,\mathrm{ms}$), with each point marked according to its density bin. The two quantities show no pronounced relation, with large scatter, and the points show no ordering by density along either axis. This is consistent with $|\alpha_{\rm DYN}|$ being set by the small-scale helical motions of the turbulence rather than by the large-scale length that fixes $\ell_{\rm mix}$.
    }
\label{fig:lmix_vs_alpha_dyn}
\end{figure}

Figure~\ref{fig:lmix_vs_alpha_dyn} compares the dynamo coefficient $|\alpha_{\rm DYN}|$ against $\ell_{\rm mix}$. The points show no clear relation, and the scatter is large. They also show no ordering by density along either axis, consistent with the non-monotonic, density-independent behavior seen in Fig.~\ref{fig:alpha_dyn_vs_rho}. $|\alpha_{\rm DYN}|$ is set by the small-scale helical motions of the turbulence and is not tied to any single large-scale length of the system. Neither $\alpha_{\rm vis}$ nor $|\alpha_{\rm DYN}|$ therefore tracks the mixing length, even though all three are measured from the same saturated flow.

\begin{figure}
    \centering
    \includegraphics[width=\linewidth]{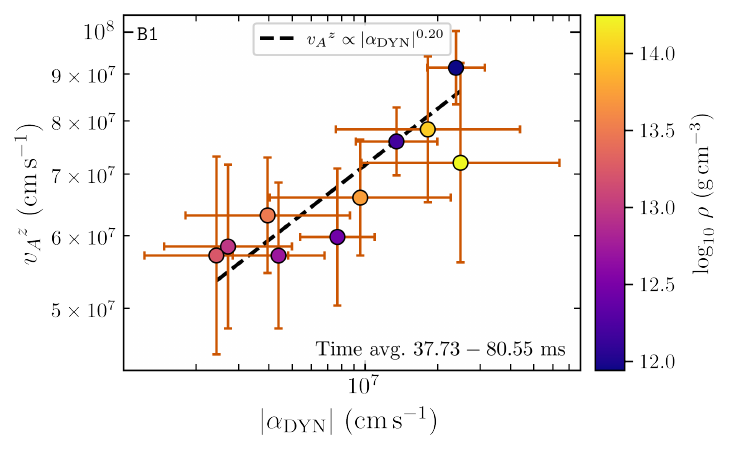}
    \caption{Vertical Alfv\'en speed $v_A{}^z$ against the dynamo coefficient $|\alpha_{\rm DYN}|$ for \texttt{B1}, time-averaged over the saturated window ($37.73-80.55\,\mathrm{ms}$), with density-weighted points. The dashed line is a weighted power-law fit, $v_A{}^z\propto|\alpha_{\rm DYN}|^{0.20}$. The Alfv\'en speed depends directly on the field strength, while $\alpha_{\rm DYN}$ depends on it only through the field-aligned fluctuation EMF, so the two are not tied together by construction. The sub-linear exponent means $v_A{}^z$ varies slowly with $|\alpha_{\rm DYN}|$, with a decade in $|\alpha_{\rm DYN}|$ spanning less than a factor of two in $v_A{}^z$.
    }
  \label{fig:alphaDYN_vs_vAz}
\end{figure}

The last dynamo comparison is between the dynamo coefficient $|\alpha_{\rm DYN}|$ and the vertical Alfv\'en speed $v_A{}^z$ (Fig.~\ref{fig:alphaDYN_vs_vAz}). The Alfv\'en speed depends directly on the field strength through $v_A{}^z=|B^z|/\sqrt{4\pi\rho}$, whereas $\alpha_{\rm DYN}$ depends on the field only indirectly, through the field-aligned part of the fluctuation EMF (Eq.~\eqref{eq:alpha_dyn_scalar}). A close relation between the two is therefore not guaranteed in advance. The weighted power-law fit gives $v_A{}^z\propto|\alpha_{\rm DYN}|^{0.20}$ for \texttt{B1}. The exponent is sub-linear, so $v_A{}^z$ varies slowly with $|\alpha_{\rm DYN}|$. A decade in $|\alpha_{\rm DYN}|$ corresponds to less than a factor of two in $v_A{}^z$. The same shallow positive scaling appears in \texttt{B2}, with $v_A{}^z\propto|\alpha_{\rm DYN}|^{0.18}$ (see Appendix~Fig.~\ref{fig:vAz_vs_alphaDYN_A2_B2_appendix}), and the exponent differs between the two models, so the scaling is not universal. Models \texttt{A1} and \texttt{A2} collapse before the turbulence saturates, so neither provides a valid measure of this relation in the saturated state.

\subsection{The mixing length and the MRI wavelength}
\label{sec:res_lmix_lambda}

\begin{figure}
\centering
\includegraphics[width=\linewidth]{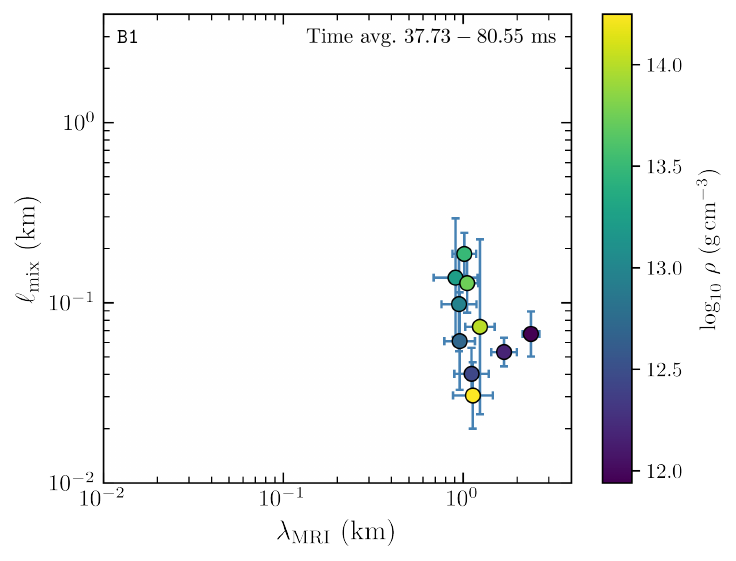}
    \caption{
    Mixing length $\ell_{\rm mix}$ against the MRI wavelength $\lambda_{\rm MRI}$ for model \texttt{B1}, time-averaged over the saturated window ($37.73-80.55\,\mathrm{ms}$) with density-weighted points. $\ell_{\rm mix}$ and $\lambda_{\rm MRI}$ are plotted to the same scale and it forms a tight vertical cluster: $\lambda_{\rm MRI}$ is concentrated near a single characteristic scale of order $1\,\mathrm{km}$ with little spread, while $\ell_{\rm mix}$ has a much larger variation across the full density range. The ratio $\ell_{\rm mix}/\lambda_{\rm MRI}$ spans $(10^{-2}-10^{-1})$ consistently across all density bins. This shows that the $\ell_{\rm mix}$-prescription is not adequately capturing the dynamics of the MRI.
   }
\label{fig:lmix_vs_lambda_mri}
\end{figure}

Figure~\ref{fig:lmix_vs_lambda_mri} plots $\ell_{\rm mix}$ directly against $\lambda_{\rm MRI}$, and the points collapse into a tight vertical cluster. $\lambda_{\rm MRI}$ has saturated to a single characteristic scale of order $1\,\mathrm{km}$ with very little spread, while $\ell_{\rm mix}$ varies by one or two orders of magnitude across the full density range. $\ell_{\rm mix}$ is therefore not a clear function of $\lambda_{\rm MRI}$. What the figure does show consistently, across all density bins, is that the ratio $\ell_{\rm mix}/\lambda_{\rm MRI}$ falls in the range $(10^{-2}-10^{-1})$. Here $\ell_{\rm mix}$ is the phenomenological transport parameter defined through $\nu_T = \ell_{\rm mix}\ c_s$. It is the effective length scale that reproduces the measured turbulent transport, and not the physical eddy scale. We therefore read the clustering in Fig.~\ref{fig:lmix_vs_lambda_mri} as a statement about this effective transport scale, not about the size of turbulent structures.

With that interpretation in mind, the time series in  Figs.~\ref{fig:time_series} and~\ref{fig:time_series_appendix} show that once the MRI saturates, $\ell_{\rm mix}$ settles to a roughly fixed fraction of $\lambda_{\rm MRI}$ rather than growing independently. As discussed in Sect.~\ref{sec:res_timeseries}, the absolute saturated value of $\lambda_{\rm MRI}$ is comparable to the resolved scale of the simulation. It is therefore the order-of-magnitude ratio $\ell_{\rm mix} \approx 0.1\,\lambda_{\rm MRI}$, rather than the absolute kilometer-scale value, that we regard as the robust and physically transferable result for subgrid modeling.

\section{Conclusion}

We have measured the effective transport coefficients of MRI-driven turbulence in differentially rotating NSs using full GRMHD \textsc{GR-Athena++} simulations. We characterized how those coefficients evolve in time, depend on density, and relate to one another.

In every model, the MRI feeds on the differential rotation, amplifying the magnetic field exponentially before saturating  as seen in Fig.~\ref{fig:B_evolution_all}. The resulting outward angular-momentum transport changes the rotation profile (see Fig.~\ref{fig:omega_profile_b1}) by spinning down the core and spinning up the bulk of each star. The consequences depend on the stellar stability of the NS. The hypermassive model \texttt{A1} loses its differential-rotation support and collapses, whereas the non-collapsing models readjust toward a more uniformly rotating equilibrium. The growth times we measure are broadly consistent with the axisymmetric results of Duez {et al.}~\cite{Duez2006Evolution}, where the discrepancies we do see are not unexpected, as our evolution takes place in three dimensions.

As seen in Fig.~\ref{fig:channel_modes_B1}, the linear MRI phase produces well-defined channel modes, seen most clearly in the alternating radial bands of the shear tensor, confirming that the instability is operating and at least partially resolved with an initial quality factor $Q=10.96$ (\texttt{B1}) at the maximum density. Their disruption ushers in the saturated turbulent state, with spatial structure we mapped through the dynamo coefficient and the mixing length (as seen in Figs.~\ref{fig:alpha_DYN_B1} and \ref{fig:lmix_B1}). The sign reversal of $\alpha_{\rm DYN}$ across the equatorial plane indicates organized helical turbulence, while the filamentary $\ell_{\rm mix}$ field reflects fully developed mixing.

The joint evolution of the four coefficients (see Figs.~\ref{fig:time_series} and \ref{fig:time_series_appendix}) is the core result. All four follow a common pattern, a sharp transient followed by a long, statistically steady plateau, and the saturated state is recovered in every model that sustains turbulence, with \texttt{A2} collapsing before the MRI can start developing, in contrast to the channel-flow structures reported by Siegel {et al.}~\cite{Siegel2013}. The source(s) of this discrepancy are unclear. The key quantitative finding is that the mixing length saturates at $\ell_{\rm mix}\approx(10^{-2}-10^{-1})\,\lambda_{\rm MRI}$, an order of magnitude below the customary $\ell_{\rm mix}\approx\lambda_{\rm MRI}$ assumption~\cite{Radice2020Calibrated}, implying that the standard closure overestimates the turbulent transport length.

The coefficients differ sharply in how they depend on density. The effective viscosity $\alpha_{\rm vis}$ follows a robust monotonic power-law decline with density in every turbulent model (Figs.~\ref{fig:alpha_vis_vs_rho} and \ref{fig:alpha_vis_vs_rho_A2_B2_appendix}), but with a model-dependent exponent, so it is structured yet non-universal, echoing the radially varying $\alpha$ of accretion discs~\cite{Penna2013VariableAlpha}. The mixing length $\ell_{\rm mix}$, by contrast, is nearly density-independent (Fig.~\ref{fig:lmix_vs_rho}), holding a fixed fraction $\ell_{\rm mix}\sim(10^{-2}-10^{-1})\,\lambda_{\rm MRI}$ throughout saturation. The two show no correlation with each other (Fig.~\ref{fig:lmix_vs_alpha_vis}): $\alpha_{\rm vis}$ spans roughly four orders of magnitude across the density range while $\ell_{\rm mix}$ stays confined to a narrow band. A single global value would therefore misrepresent $\alpha_{\rm vis}$, while imposing a density dependence would not be appropriate for $\ell_{\rm mix}$.

The dynamo coefficient $\alpha_{\rm DYN}$ occupies a distinct role among the diagnostics. Its clear antisymmetry across the equatorial plane (Fig.~\ref{fig:alpha_DYN_B1}) confirms that the turbulence carries organized helicity driven by the Coriolis force, the hallmark of a coherent mean-field $\alpha$-effect. In saturation, however, $|\alpha_{\rm DYN}|$ shows no clean dependence on density (Fig.~\ref{fig:alpha_dyn_vs_rho}) and no correlation with $\ell_{\rm mix}$ (Fig.~\ref{fig:lmix_vs_alpha_dyn}). Its only statistically meaningful trend is a shallow positive scaling with the vertical Alfv\'{e}n speed $v_A{}^z\propto|\alpha_{\rm DYN}|^{0.20}$ in \texttt{B1} and $|\alpha_{\rm DYN}|^{0.18}$ in \texttt{B2} (Figs.~\ref{fig:alphaDYN_vs_vAz} and \ref{fig:vAz_vs_alphaDYN_A2_B2_appendix}), with a sub-linear, model-dependent exponent. This behavior is consistent with $\alpha_{\rm DYN}$ being set by the small-scale helical structure of the turbulence rather than any single large-scale property of the system.

The one tight relation among all the diagnostics is between the two length scales. Plotting $\ell_{\rm mix}$ directly against $\lambda_{\rm MRI}$ (Fig.~\ref{fig:lmix_vs_lambda_mri}) shows that $\lambda_{\rm MRI}$ saturates to a single macroscopic scale set by the global properties of the star, while $\ell_{\rm mix}$ varies by one or two orders of magnitude across the density range. The ratio $\ell_{\rm mix}/\lambda_{\rm MRI}$ nevertheless falls consistently in $(10^{-2}-10^{-1})$ across all density bins and all turbulent models. Given the bulk properties of the system as encapsulated by $\lambda_{\rm MRI}$ and $v_A$, we can at best provide order-of-magnitude estimates of the individual transport coefficients rather than precise predictions. The ratio $\ell_{\rm mix}\approx0.1\,\lambda_{\rm MRI}$ is therefore the one physically transferable result for subgrid modeling, lying an order of magnitude below the standard assumption $\ell_{\rm mix}\approx\lambda_{\rm MRI}$~\cite{Radice2020Calibrated}.

Several limitations of this study should be kept in mind. Our models adopt a $\Gamma=2$ ideal-gas equation of state and the $j$-constant rotation law, both of which are idealizations of realistic post-merger remnants. The $j$-constant profile differs substantially from the rotation profiles measured in merger simulations~\cite{Hanauske2017,Uryu2017RotationLaws}, and we work in the ideal-MHD limit without neutrino radiation or finite-temperature effects, both of which can alter field amplification and turbulent structure in realistic remnants~\cite{Kiuchi2024Review}. Our simulations are also initialized with a purely poloidal seed field, whereas post-merger remnants carry a more complex field geometry that could affect the early channel structure and saturation values~\cite{AguileraMiret2024DelayedJet,Bamber2025Topology,Rainho2025Topology}. Finally, our quality factors are $Q_{\rm MRI}\sim 7-100$ at $t=0$, close to the minimum for resolving the linear MRI phase in the least favorable case. Furthermore, higher-resolution runs are needed to test whether the saturated transport coefficients, particularly $\ell_{\rm mix}/\lambda_{\rm MRI}$, are converged. Future work will extend these measurements to more realistic differential rotation profiles, motivated by binary merger simulations and to finite-temperature equations of state that better represent post merger remnants. Together these extensions will test the robustness of the calibrated transport coefficients and their suitability for BNS merger remnants. 

\begin{acknowledgments}
 DR acknowledges support from the U.S.~Department of Energy, Office of Science, Division of Nuclear Physics under Award Number(s) DE-SC0024388, and from the National Science Foundation under Grants PHY-2020275, PHY-2116686, PHY-2407681, PHY-2512802, and PHY-2621752.
Computing resources were provided by the Texas Advanced Computing Center (TACC) at the University of Texas at Austin through the Leadership Resource Allocation (LRAC) Grant AST-2108467. The Frontera and Vista computing projects are made possible by National Science Foundation award OAC-1818253. Data analysis and postprocessing was performed using the Roar Colab supercomputer at the Institute for Computational and Data Science of the Pennsylvania State University.
\end{acknowledgments}

\appendix

\section{Additional Models}
\label{appendix}

We present the transport coefficient time evolution along with relevant density scaling plots for \texttt{A1}, \texttt{A2}, and \texttt{B2} that were not featured in the main text.

\begin{figure*}
\centering
\includegraphics[width=\linewidth]{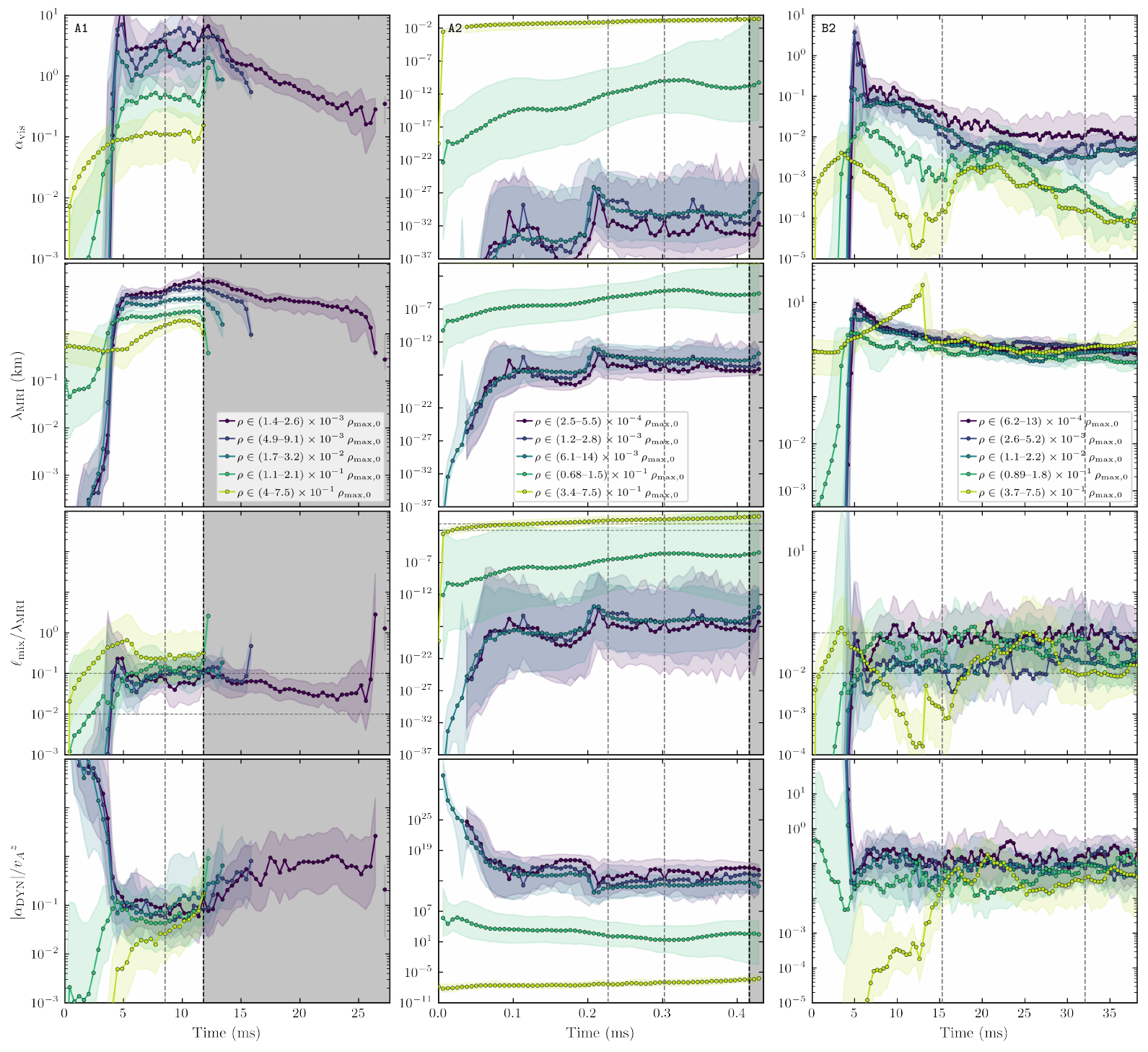}
    \caption{Time evolution of the four transport coefficients ($\alpha_{\rm vis}$, $\lambda_{\rm MRI}$,
    $\ell_{\rm mix}/\lambda_{\rm MRI}$, and $|\alpha_{\rm DYN}|/v_A^z$ rows) for \texttt{A1}, \texttt{B2}, and \texttt{A2} (columns), with curves weighted by density. Vertical dashed lines mark the saturated averaging window and the grey shading marks the post-collapse region in \texttt{A1}. \texttt{B2} reproduces the similar quasi-steady turbulent state as \texttt{B1} (Fig.~\ref{fig:time_series}). \texttt{A1} reaches the same plateau before collapsing. \texttt{A2} forms a BH within $\lesssim0.5\,\mathrm{ms}$, before MRI turbulence develops, and shows no saturated structure.}
\label{fig:time_series_appendix}
\end{figure*}
Figure~\ref{fig:time_series_appendix} shows the time evolution of all four transport coefficients for these models. \texttt{B2} reaches the same quasi-steady turbulent plateau as \texttt{B1}, confirming that MRI saturation is reproducible across the non-collapsing models. \texttt{A1} achieves a similar plateau before gravitational collapse terminates the channel mode destruction. \texttt{A2} collapses to a black hole within $\lesssim 0.5\,\mathrm{ms}$, before MRI turbulence can develop, and shows no saturated structure in any coefficient.

\begin{figure*}
\centering
\includegraphics[width=\linewidth]{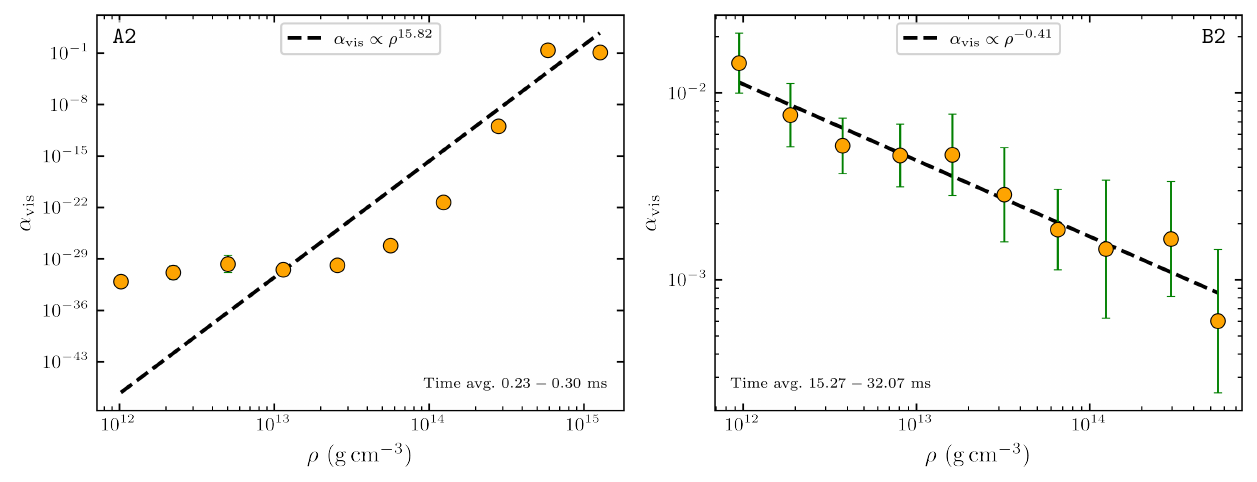}
    \caption{$\alpha_{\rm vis}$ vs $\rho$ plot similar to Fig.~\ref{fig:alpha_vis_vs_rho}, for \texttt{A2} (left) and \texttt{B2} (right). \texttt{B2} ($15.27-32.07\,\mathrm{ms}$) reproduces the same monotonic density dependence, $\alpha_{\rm vis}\propto\rho^{-0.41}$, supporting the generality of the trend across models. However, \texttt{A2} ($0.23-0.30\,\mathrm{ms}$) collapses before MRI-driven turbulence develops and never establishes a meaningful   $\alpha_{\rm vis}$; its values are vanishingly small and unstructured, and the steep fitted slope is an artifact of the collapse, not a physical density relation.}
\label{fig:alpha_vis_vs_rho_A2_B2_appendix}
\end{figure*}
Figure~\ref{fig:alpha_vis_vs_rho_A2_B2_appendix} compares $\alpha_{\rm vis}$ as a function of density for \texttt{A2} and \texttt{B2}. \texttt{B2} reproduces the monotonic power-law trend $\alpha_{\rm vis} \propto \rho^{-0.41}$ found in \texttt{B1}, supporting the robustness of this density dependence across the non-collapsing models. For \texttt{A2}, no physical density relation is established. The fitted slope is an artifact of the pre-collapse state, not a signature of MRI-driven turbulence.

\begin{figure*}
\centering
    \includegraphics[width=\linewidth]{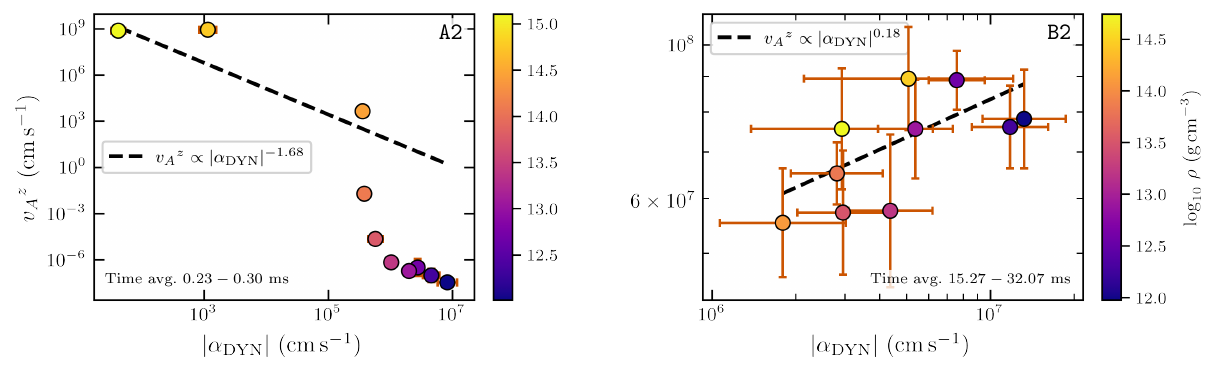}
        \caption{Similar to Fig.~\ref{fig:alphaDYN_vs_vAz}, this is the vertical Alfv\'en speed $v_A{}^z$ against the dynamo coefficient $|\alpha_{\rm DYN}|$ for \texttt{A2} (left) and \texttt{B2} (right). \texttt{B2} ($15.27-32.07\,\mathrm{ms}$) recovers the same positive, shallow correlation seen in \texttt{B1} ($v_A{}^z\propto|\alpha_{\rm DYN}|^{0.18}$), consistent with both quantities tracking the magnetic field in the saturated state. \texttt{A2} ($0.23-0.30\,\mathrm{ms}$) collapses to a BH before MRI saturation, with no developed turbulent state, it does not provide a valid measure of the relation.}
    \label{fig:vAz_vs_alphaDYN_A2_B2_appendix}
\end{figure*}
Similarly, Figure~\ref{fig:vAz_vs_alphaDYN_A2_B2_appendix} shows the correlation between the vertical Alfv\'en speed $v_A{}^z$ and the dynamo coefficient $|\alpha_{\rm DYN}|$ for the same two models. \texttt{B2} recovers the shallow positive correlation seen in \texttt{B1}, with both quantities tracking the saturated magnetic field strength. \texttt{A2} does not provide a valid measure of this relation, as it never reaches a turbulent steady state.

\bibliography{ref}

\end{document}